\documentclass[10pt,journal,final,twocolumn]{IEEEtran}
\usepackage{epsfig,graphicx,subfigure,psfrag,amsmath,cases,bm}
\usepackage{latexsym,amssymb,algorithm,mathtools}
\usepackage{algorithmic}
\usepackage{color}
\usepackage{url}
\usepackage{scrtime}
\usepackage{stfloats}
\usepackage{tablefootnote}
\usepackage{cite}
\usepackage{amsfonts,mathabx}
\usepackage{textcomp}
\usepackage{xcolor,capt-of}
\usepackage{multirow}
\usepackage{amsthm}
\usepackage{subfigure}
\usepackage{makecell}
\usepackage{pifont}
\usepackage{amssymb}
\allowdisplaybreaks[3]

\newtheorem{remark}{Remark}
\newtheorem{lemma}{Lemma}

\newtheorem{T-Prob}{Transformed Problem}

\DeclareMathOperator{\maxo}{maximize}
\DeclareMathOperator{\mino}{minimize}

\DeclareMathOperator{\subto}{subject\hspace*{2mm}to}

\newcommand{\myoverline}[1]{\overline{\overline{#1}}}
\newcommand{\mywidetilde}[1]{\widetilde{\widetilde{#1}}}
\newcommand{\ul}[1]{\underline{#1}}
\newcommand{\ui}{^{(i)}}
\newcommand{\uu}{^{(1)}}
\newcommand{\ii}{^{(2)}}

\begin{document} 

\title{Adaptive Sensing Performance Design for Enhancing Secure Communication in Networked ISAC Systems}

\author{Yiming Xu, \textit{Graduate Student Member, IEEE,}  Dongfang Xu, \textit{Member, IEEE,} \\ Shenghui Song, \textit{Senior Member, IEEE,} and Dusit Niyato, \textit{Fellow, IEEE}
\thanks{Yiming Xu and Shenghui Song are with Department of Electronic and Computer Engineering, The Hong Kong University of Science and Technology, Hong Kong, China (email: yxuds@connect.ust.hk, eeshsong@ust.hk); Dongfang Xu is with Division of Integrative Systems and Design, The Hong Kong University of Science and Technology, Hong Kong, China (email: eedxu@ust.hk); 
Dusit Niyato is with College of Computing and Data Science, Nanyang Technological University, Singapore (e-mail: dniyato@ntu.edu.sg).
}
}

\maketitle
\begin{abstract}
The channel state information (CSI) of an eavesdropper is crucial for physical layer security (PLS) design, but it is difficult to obtain due to the passive and non-cooperative nature of the eavesdropper. To this end, integrated sensing and communication (ISAC) offers a novel solution by estimating the CSI of the eavesdropper based on sensing information.
However, existing studies normally impose explicit and fixed sensing performance requirement without considering the varying communication conditions, which hinders the system from fully exploiting the synergy between sensing and communication. To address this issue, this paper proposes sensing-enhanced secure communication with adaptive sensing performance. Specifically, we formulate the sensing performance implicitly in the information leakage rate and adaptively optimize it for the minimization of the power
consumption, offering enhanced flexibility and adaptability in sensing performance.
We consider both centralized and decentralized designs to thoroughly investigate the impact of network structure on system performance and complexity. Specifically, we devise a block coordinate descent (BCD)-based method for centralized design. For decentralized design, we develop an optimization framework based on consensus alternating direction method of multipliers (ADMM) to reduce complexity and information exchange overhead.
Experimental results demonstrate the advantage of the proposed implicit sensing performance requirement design due to its capability to adaptively adjust the sensing performance to enhance the system performance for varying system configurations.

\end{abstract}


\section{Introduction}
The unshielded channels of wireless communication make it vulnerable to eavesdropping, posing significant risks to data confidentiality. As an effective alternative to traditional cryptographic methods, physical layer security (PLS) leverages the unique properties of wireless channels to provide secure communication by techniques such as beamforming and artificial noise injection \cite{7470273, 10183798, 11018844}. For that purpose, acquiring channel state information (CSI) of the eavesdropper is essential but remains a challenging task due to the passive and non-cooperative nature of the eavesdropper \cite{10608156, 10622357}.
\par
Fortunately, the integrated sensing and communication (ISAC) technique provides a new paradigm for PLS design where the CSI of the eavesdropper can be estimated through sensing \cite{9199556,9933849,10870062}. Specifically, Su \textit{et al.} \cite{9199556} estimated the CSI of the eavesdropper based on angle and angle uncertainty provided by sensing. Then, the sum of the eavesdropper's signal-to-noise ratio (SNR) within the angle uncertainty set was minimized. 
Xu \textit{et al.} \cite{9933849} constructed the CSI error model of the eavesdropper using its angle and distance, and derived the CSI error bound caused by the estimation uncertainty. Afterwards, a sequential beam scanning method was designed for robust and secure communication. However, the above studies did not consider the practical process of acquiring sensing information.
\par
To this end, a two-stage design has been proposed in the literature, where sensing is performed in the first stage to obtain the CSI of the eavesdropper, based on which secure communication is achieved in the second stage \cite{10227884, 9645576, 10349846}. In particular, Su \textit{et al.} \cite{10227884} proposed to use the omnidirectional waveform for eavesdropper detection. Then, the weighted sum of the scaled Cram\'{e}r-Rao bound (CRB) and secrecy rate is optimized at the second stage based on the estimated angle of the eavesdropper. Wang \textit{et al.} \cite{9645576} proposed to first optimize the sensing performance measured by the difference between mainlobe and sidelobe, and then maximize the signal-to-interference-plus-noise ratio (SINR) of the received uplink signals while restricting the SINR of the eavesdropper.
To further improve the performance of the two-stage design, Xu \textit{et al.} \cite{10349846} investigated the variable-length two-stage design, where the duration ratio of the two stages is optimized, providing additional degree of freedom (DoF) for system design. In the first stage, CRB is minimized to achieve the best sensing estimation of the eavesdropper, based on which the second stage performs the secure information transmission. However, in the above design, the sensing performance is explicitly imposed in the first stage without considering the communication requirement in the second stage. Furthermore, varying communication requirements and channel conditions demand adaptable sensing performance control, which has yet to be thoroughly investigated in existing studies. As a result, the synergy between sensing and communication for PLS provisioning is not fully exploited.


\par
To tackle the above-mentioned issue, in this paper, we investigate the design of sensing-enhanced secure communication in networked ISAC systems with adaptive sensing performance. Different from existing methods that explicitly impose sensing performance requirement, we implicitly formulate the sensing performance in information leakage rate without imposing any restrictions on the sensing performance.
Compared with the explicit sensing performance requirement, the proposed design can adaptively determine the preferred sensing performance to boost the system performance. Implicit sensing performance optimization allows for greater flexibility in sensing performance, expanding the feasible region of the optimization problem and thereby potentially enhancing overall system performance. 
We consider both centralized and decentralized designs to thoroughly investigate the impact of network structure on system performance and complexity.
In particular, the total power consumption of all base stations (BSs) is minimized under the constraints of achievable rate and information leakage rate, where the sensing performance is optimized to meet the requirement of secure communication.
\par
We examine both centralized and decentralized designs to conduct a comprehensive exploration of the network structure. For that purpose, we first derive the CRB for both schemes and then develop the associated optimization framework.
The main challenge for the centralized design lies in the implicit involvement of sensing performance in the information leakage rate. To this end, we exploit block coordinate descent (BCD) \cite{tseng2001convergence} theory and develop a computationally-efficient optimization method to address this issue. To further reduce the computational complexity and information exchange overhead required for BS coordination, we then introduce a decentralized design, where each BS uses only its local information to collaboratively solve the problem. However, the involvement of the inter-BS interference in the achievable rate and information leakage rate complicates the decomposition of the problem for decentralized design. To this end, we develop a consensus alternating direction method of multipliers (ADMM) \cite{boyd2011distributed} to handle the problem.
\par
The main contributions of this paper are summarized as follows.
\begin{itemize}
\item We investigate sensing-enhanced secure communication with an implicit sensing performance requirement. Compared to existing methods that explicitly impose sensing requirements, the proposed design enables adaptive optimization of sensing performance to maximize system efficiency.
\item For centralized design, we first derive the CRB for centralized sensing and then develop a BCD-based centralized optimization framework. The centralized design fully exploits the system performance provided by the networked structure by directly utilizing the CSI and original sensing echoes from all BSs.
\item For decentralized design, we first derive the CRB of decentralized sensing. Then, we develop a decentralized optimization framework based on consensus ADMM, in which the sets of global and local variables are maintained for information sharing among BSs. Remarkably, decentralized
design offers a substantial reduction in computational complexity and information exchange overhead compared to its centralized counterpart.
\item Numerical results reveal the existence of an optimal sensing performance that maximizes system performance. This indicates that explicit sensing performance requirements are unnecessary and undermine system performance. Compared to the existing methods, the proposed design benefits from its ability to dynamically adapt sensing performance to accommodate changing system configurations.
\end{itemize}
\par
The rest of the paper is organized as follows. In Section II, we present the system model and problem formulation. Section III presents the centralized design, and Section IV shows the decentralized design. Numerical results are shown in Section V. Finally, Section VI concludes the paper.



\begin{figure*}[t]
\centering
\includegraphics[width=6.6in]{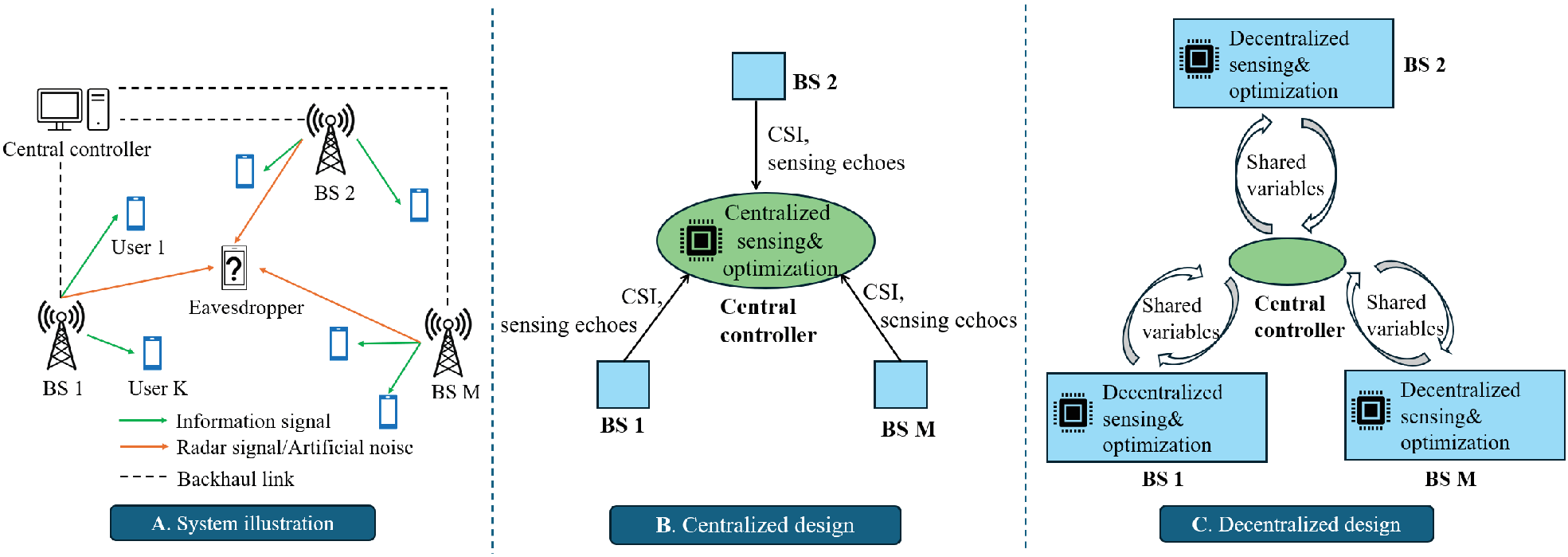}
\caption{An overview of the sensing-enhanced secure communication in a networked ISAC system. \textit{Part A} illustrates the architecture of the considered system, where multiple BSs perform collaborative sensing to localize the eavesdropper and serve the associated users. The BSs are connected to a central controller for information exchange and coordination. 
\textit{Part B} presents the topological structure of the centralized design. 
The central controller collects the CSI and sensing echoes from all BSs and performs the centralized sensing and optimization. \textit{Part C} presents the topological structure of the decentralized design. BSs perform decentralized sensing and optimization in parallel, with information shared with other BSs via the central controller.}
\label{figure:system_illus}
\end{figure*}
\section{System Model and Problem Formulation}
In this paper, we consider a networked ISAC system consisting of one single-antenna eavesdropper, $M$ BSs, each serving $K$ single-antenna users, as shown in Fig. \ref{figure:system_illus} \textit{Part A}.\footnote{The association between the BSs and users can be resolved using the existing user association method \cite{10602493,10207991}. Without loss of generality, we assume that each BS serves the same number of users. We note that the proposed algorithm is fully capable of handling a more general scenario where each BS serves a different number of users.} Each BS is equipped with a uniform linear array (ULA) comprising $N$ antennas and connected to a central controller via backhaul links \cite{10726912}.
\subsection{Two-stage protocol}
\begin{figure}[t]
\centering
\includegraphics[width=3.4in]{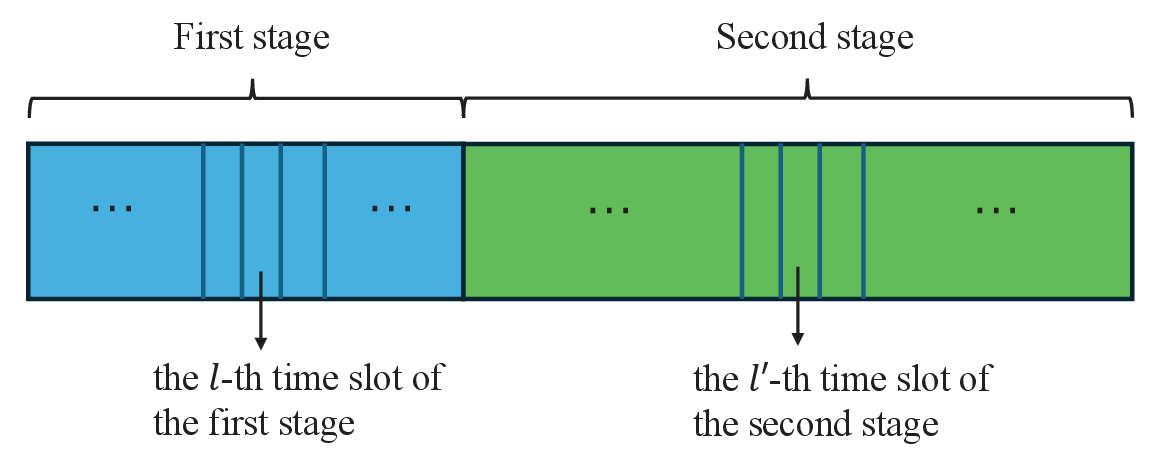}
\caption{An illustration of the two-stage protocol of the frame. Each frame is divided into two stages, each of which consists of multiple time slots. In the first stage, the BSs simultaneously conduct sensing and communication. In the second stage, the secure information transmission is carried out, utilizing the refined CSI obtained from the sensing results in the first stage.}
\label{figure:time_frame}
\end{figure}

We adopt a two-stage protocol where each frame is divided into two stages, as illustrated in Fig.~\ref{figure:time_frame}.
In the first stage, the BSs transmit the dual-functional radar-communication (DFRC) signal for both sensing and secure communication purposes based on the coarse location estimation of the eavesdropper from the previous frame. In this stage, the sensing signal also serves as artificial noise for secure communication \cite{9199556}.\footnote{Due to the change of the channel conditions caused by the dynamic environment, the system can only obtain a coarse estimation of the eavesdropper's location based on the sensing result in the last frame and the state evolution function \cite{10345500}.}
At the end of the first stage, an accurate location estimation is obtained using the received echoes, based on which the eavesdropper's CSI is estimated. The estimated CSI is then utilized in the second stage, where the BSs transmit information signals and artificial noise for secure communication. The normalized duration of the $i$-th stage is denoted by $\tau_i$, $i \in \mathcal{I} \overset{\triangle}{=} \{1,2\}$, where $\tau_1 +  \tau_2 = 1$.

\subsection{Centralized and Decentralized Design}
For centralized design, the central controller collects the received sensing echoes and the CSI estimated at all BSs. 
The computation for sensing and solving the transmit beamforming of all BSs is carried out at the central controller, as shown in Fig. \ref{figure:system_illus} \textit{Part B}. For decentralized design, each BS first performs sensing estimation based on its own received sensing echoes. Then, the BS reports its estimation to the central controller for joint location estimation \cite{554211}. Each BS designs its transmit beamforming based on its local CSI and the eavesdropper's CSI, which is obtained through joint location estimation, as shown in Fig. \ref{figure:system_illus} \textit{Part C}.
\subsection{Signal Model}
For notational simplicity, we define sets $\mathcal{M} \overset{\triangle}{=} \left\{ 1, \dots, M\right\}$ and $\mathcal{K} \overset{\triangle}{=} \left\{ 1, \dots, K\right\}$ to present the indices of the BSs and users, respectively. We use the tuple $(m,k)$ to index the $k$-th user served by the $m$-th BS. 
The transmitted signal from the $m$-th BS in the $l$-th time slot during the $i$-th stage is given by
\begin{align}
\mathbf{x}^{(i)}_m[l] = \underset{k \in \mathcal{K}}{\sum} \mathbf{w}_{m,k}\ui s_{m,k}\ui[l] + \mathbf{r}_m\ui[l],
\end{align}
where $\mathbf{w}_{m,k}\ui \in \mathbb{C}^{N \times 1}$ denotes the beamforming vector transmitted from the $m$-th BS to the $(m,k)$-th user in the $i$-th stage. 
$s_{m,k}[l] \sim \mathcal{CN}(0,1)$ denotes the information signal for the $(m,k)$-th user. The signals for different users are assumed to be independent. Vector $\mathbf{r}_m\ui \in \mathbb{C}^{N \times 1}$ is the dedicated sensing signals and is independent of the communication signals.
The covariance matrix of $\mathbf{r}_m\ui[l]$ is denoted by $\mathbf{R}_m\ui \in \mathbb{C}^{N \times N}$. As a result, the covariance matrix of the transmitted signal $\mathbf{x}_m\ui$ is given by
\begin{align}
\mathbf{S}_m\ui = \underset{k \in \mathcal{K}}{\sum} \mathbf{w}_{m,k}\ui (\mathbf{w}_{m,k}\ui)^H + \mathbf{R}_m\ui.
\end{align}
\subsubsection{Received Signals at Users}
\par
The received signal at the $(m,k)$-th user is given by
\begin{align} \label{comm_signal}
y_{m,k}\ui[l] = & \underset{m' \in \mathcal{M}}{\sum} \mathbf{h}_{m',m,k}^H\mathbf{x}_{m'}\ui[l] + n_{m,k}[l],
\end{align}
where $\mathbf{h}_{m',m,k} \in \mathbb{C}^{N \times 1}$ denotes the channel between the $m'$-th BS and the $(m,k)$-th user, and $n_{m,k}[l] \sim \mathcal{CN}(0,\sigma_{m,k}^2)$ represents the additive white Gaussian noise (AWGN) at the $(m,k)$-th user. 
We rewrite \eqref{comm_signal} as
\begin{align}
y_{m,k}\ui[l] =\hspace*{1mm}& \mathbf{h}_{m,m,k}^H\mathbf{w}_{m,k}\ui s_{m,k}\ui[l] + \hspace*{-2mm}
\underset{k' \in \mathcal{K} \setminus \{k\} }{\sum} \mathbf{h}_{m,m,k}^H\mathbf{w}_{m,k'}\ui s_{m,k'}\ui[l] \notag \\
& + \underset{m' \in \mathcal{M} \setminus \{m\}}{\sum} \underset{k' \in \mathcal{K} }{\sum} \mathbf{h}_{m',m,k}^H\mathbf{w}_{m',k'}\ui s_{m',k'}\ui[l] \notag \\
& + \underset{m' \in \mathcal{M} }{\sum} \mathbf{h}_{m',m,k}^H \mathbf{r}_{m'}\ui[l]  + n_{m,k}[l].
\end{align}
Thus, the SINR of the $(m,k)$-th user in the $i$-th stage is given by \eqref{sinr}, shown at the top of the next page. As a result, the achievable rate is given by
\setcounter{equation}{5}
\begin{align} \label{achievable_rate}
R_{m,k}\ui = \log_2(1 + \gamma_{m,k}\ui).
\end{align}
\begin{figure*}[t]
\setcounter{equation}{4}
\begin{align} \label{sinr}
\gamma_{m,k}\ui = \frac{ \left| \mathbf{h}_{m,m,k}^H \mathbf{w}_{m,k}\ui \right|^2 }{\underset{k' \in \mathcal{K} \setminus \{k\} }{\sum}  \left| \mathbf{h}_{m,m,k}^H \mathbf{w}_{m,k'}\ui \right|^2 + \underset{m' \in \mathcal{M} \setminus \{m\}}{\sum} \underset{k' \in \mathcal{K} }{\sum} \left| \mathbf{h}_{m',m,k}^H\mathbf{w}_{m',k'}\ui \right|^2 + \underset{m'\in \mathcal{M}}{\sum} \mathbf{h}_{m',m,k}^H \mathbf{R}_{m'}\ui \mathbf{h}_{m',m,k} + \sigma_{m,k}^2}.
\end{align}
\hrule
\setcounter{equation}{6}
\end{figure*}
\subsubsection{Received Signals at the Eavesdropper}
In this paper, we assume that the eavesdropper has already been detected, which can be achieved by exploiting the local oscillator leakage power \cite{6288501}.
The received signal at the eavesdropper in the $i$-th stage is given by
\begin{align}
y_{\text{e}}\ui[l] & = \underset{m \in \mathcal{M}}{\sum} \mathbf{g}_m^H \mathbf{x}_m\ui[l] + n_{\text{e}}[l] \notag \\
& = \underset{k \in \mathcal{K}}{\sum} \underset{m \in \mathcal{M}}{\sum} \mathbf{g}_m^H \mathbf{w}\ui_{m,k} s_{m,k}\ui[l] + \underset{m \in \mathcal{M}}{\sum} \mathbf{g}_m^H \mathbf{r}_m\ui[l] + n_e[l],
\end{align}
where $n_{\text{e}}[l] \sim \mathcal{CN}(0, \sigma_{\text{e}}^2)$ denotes the AWGN. Vector $\mathbf{g}_m \in \mathbb{C}^{N \times 1}$ represents the channel between the $m$-th BS and the eavesdropper, given by
\begin{align} \label{sensing_channel}
\mathbf{g}_m = \alpha_m \left( \sqrt{\frac{\kappa}{1+\kappa}} \mathbf{a}(\theta_m) + \sqrt{\frac{1}{1+\kappa}}\widetilde{\mathbf{g}}_m \right),
\end{align}
where $\theta_m$ denotes the angle of departure (AoD) of the eavesdropper with respect to the $m$-th BS, $\alpha_m$ represents the path gain, and $\kappa$ is the Rician factor. $\mathbf{a}(\theta_m) \in \mathbb{C}^{N \times 1}$ is the steering vector from the $m$-th BS to the eavesdropper, given by
\begin{align}
\mathbf{a}(\theta_m) = [1, e^{j\pi\cos\theta_m}, \dots, e^{j\pi (N-1) \cos\theta_m}]^T.
\end{align}
Vector $\widetilde{\mathbf{g}}_m$ denotes the fading channel caused by the scatterers, with a bounded norm $\beta_{\text{NLoS},m}$ \cite{9933849, 10945425}, given by
\begin{align}
\| \widetilde{\mathbf{g}}_m \| \leq \beta_{\text{NLoS},m}.
\end{align}
Assuming the worst case scenario where the eavesdropper can eliminate the multiuser interference \cite{9133130}, the leakage rate at the eavesdropper for wiretapping the information of the $(m,k)$-th user is given by
\begin{align}
\widehat{R}_{m,k}^{(i)} = \log_2 \left( 1 + \frac{ | \mathbf{g}_m^H \mathbf{w}_{m,k}\ui |^2}{ \underset{m \in \mathcal{M}}{\sum} \mathbf{g}_m^H \mathbf{R}_m\ui \mathbf{g}_m + \sigma_{\text{e}}^2 } \right).
\end{align}
\subsection{CSI Uncertainty with Sensing Estimation Errors}
Sensing estimation errors inevitably lead to uncertainty in the eavesdropper's CSI, i.e., $\mathbf{g}_m$ in \eqref{sensing_channel} \cite{zhao2025generative}. In this subsection, we model the CSI error to facilitate the robust design for secure communication.
\par
We define the location uncertainty by $\Delta \mathbf{p} = [\Delta p_1, \Delta p_2]^T = \mathbf{p} -\overline{\mathbf{p}}$, where $\mathbf{p} \overset{\triangle}{=} [p_1, p_2]^T$ and $\overline{\mathbf{p}}$ denote the true and estimated value of the 2-dimensional Cartesian coordinates of the eavesdropper, respectively.
According to the three-sigma rule \cite{10781436, 10812799}, the location uncertainty at the $i$-th stage can be approximated by $| \Delta p_j | \leq 3 \sqrt{[\mathbf{Q}\ui]_{j,j}}$, $j \in \{1,2\}$. $\mathbf{Q}\ui$ denotes the covariance matrix of the location estimation utilized in the $i$-th stage. 
\begin{remark}
$\mathbf{Q}\uu$ is obtained at the beginning of a frame based on the sensing estimation in the last frame and the state prediction. $\mathbf{Q}\ii$ results from the sensing estimation in the first stage and is measured by CRB, which depends on information beamforming $\mathbf{w}\uu_{m,k}$ and radar covariance matrix $\mathbf{R}\uu_m$. We derive the detailed expression of $\mathbf{Q}\ii$ in Section \ref{centralized_sensing} for centralized sensing and Section \ref{decentralized_sensing} for decentralized sensing, respectively.
\end{remark}
In the following, we first derive the uncertainty of AoD $\theta_m$, and the result is then used to derive the uncertainty of the CSI. The uncertainty of the AoD $\theta_m$ is modeled as
\begin{align}
& \theta_m = \overline{\theta}_m + \Delta \theta_m, \notag \\
& \Delta \theta_m \in \Omega_{\Delta \theta_m}\ui \overset{\triangle}{=} \left\{ \Delta \theta_m: |\Delta \theta_m| \leq \beta_{\theta_m}\ui \right\},
\end{align}
where $\theta_m$ and $\overline{\theta}_m$ denote the true and estimated values of the AoD, respectively. $\Omega_{\Delta \theta_m}\ui$ is the uncertainty set, with $\beta_{\theta_m}\ui = \frac{\| \Delta \mathbf{p} \|}{\overline{d}_m} = \frac{3\sqrt{\mathrm{Tr}\left( \mathbf{Q}\ui \right)}}{\overline{d}_m}$, where $\overline{d}_m = \| \mathbf{q}_m - \overline{\mathbf{p}} \|$. Here, $\mathbf{q}_m$ denotes the location of the $m$-th BS. 
\par
Next, we derive the CSI uncertainty. The uncertainty model of $\mathbf{g}_m$ is given by
\begin{align} \label{channel_g}
\mathbf{g}_m = \overline{\mathbf{g}}_m + \Delta \mathbf{g}_m,
\end{align}
where $\mathbf{g}_m$ and $\overline{\mathbf{g}}_m$ denote the true value and the estimated value, respectively. We adopt the widely applied bounded uncertainty model to capture the CSI error $\Delta \mathbf{g}_m$ \cite{wang2009worst}. The norm of $\Delta \mathbf{g}_m$ is calculated by
\begin{align}
\| \mathbf{\Delta \mathbf{g}_m} \| = & \| \mathbf{g}_m - \overline{\mathbf{g}}_m \| \notag \\
 \leq & \alpha_m \sqrt{\frac{\kappa}{1+\kappa}} \| \mathbf{a}(\theta_m + \Delta \theta_m) - \mathbf{a}(\theta_m) \| \notag \\
& + \alpha_m \sqrt{\frac{1}{1+\kappa}} \beta_{\text{NLoS},m}.
\end{align}
Following the similar derivation in \cite{10375133}, we have
\begin{align}
& \hspace*{-6mm}\| \mathbf{a}(\theta_m + \Delta \theta_m) - \mathbf{a}(\theta_m) \|  \notag \\
\leq & \sqrt{\frac{N(N-1)(2N-1)}{6}} \pi \sin \overline{\theta}_m \beta_{\theta_m}\ui,
\end{align}
which represents the error bound of the steering vector with respect to the eavesdropper caused by the location estimation error.
As a result, the uncertainty set of $\mathbf{g}_m$ in the $i$-th stage is given by
\begin{align}
\Omega_{\Delta \mathbf{g}_m}\ui \overset{\triangle}{=} \left\{ \Delta \mathbf{g}_m : | \Delta \mathbf{g}_m | \leq \beta_{\mathbf{g}_m}\ui \right\},
\end{align}
where 
\begin{align} \label{beta_gm}
\beta_{\mathbf{g}_m}\ui = & \hspace{1mm} \alpha_m \beta_{\text{NLoS},m} \sqrt{\frac{1}{1+\kappa}} \notag \\
& \hspace*{-10mm} + \frac{3 \pi \alpha_m \sin \overline{\theta}_m }{\overline{d}_m} \sqrt{\frac{\kappa}{1+\kappa}} \sqrt{\frac{N(N-1)(2N-1)}{6}} \sqrt{\mathrm{Tr}\left( \mathbf{Q}\ui \right)}.
\end{align}
\par
The worst-case leakage rate of the 
$(m,k)$-th user in the presence of the CSI error is $\underset{i \in \mathcal{I}}{\sum} \underset{\substack{ \Delta \mathbf{g}_m \in \Omega_{\Delta \mathbf{g}_m}^{(i)} }}{\max} \tau_i \widehat{R}_{m,k}\ui$.
\begin{remark}
There is an interdependence between the sensing accuracy of the first stage and the secure communication rate of the second stage, which jointly affects system performance. Specifically, low sensing accuracy results in significant CSI estimation errors of the eavesdropper, leading to a low secure communication rate in the second stage and degrading overall system performance \cite{10781436}. Conversely, achieving
extremely high sensing accuracy provides limited improvements to secure communication while incurring significant additional power consumption. Hence, there exists a preferred sensing performance threshold that maximizes the system performance. The explicit sensing performance requirement fails to account for this balance. To address this issue, we propose an adaptive sensing performance design that dynamically balances this trade-off to improve overall efficiency.
\end{remark}

\subsection{Problem Formulation}
In this paper, we aim to minimize the total power consumption of all BSs while guaranteeing desired achievable rate of legitimate users and maximum information leakage tolerance to the eavesdropper. The optimization problem is formulated as
\begin{align} \label{main_prob}
\underset{\substack{ \mathbf{w}_{m,k}^{(i)}, \mathbf{R}_m^{(i)} }}{\mino} & \hspace*{3mm} \underset{m \in \mathcal{M}}{\sum} P_m \notag\\
\subto & \hspace*{3mm} \mbox{C1} \mbox{:} \hspace{1mm} P_m \leq P_{\text{max}}, \hspace{1mm}\forall m, \notag \\
& \hspace*{3mm} \mbox{C2} \mbox{:} \hspace{1mm} \underset{i \in \mathcal{I}}{\sum} \tau_i R_{m,k}\ui \geq R_{\text{info},m,k},\hspace{1mm}\forall m, \hspace{1mm}\forall k, \notag \\
& \hspace*{3mm} \mbox{C3} \mbox{:} \hspace{1mm} \underset{i \in \mathcal{I}}{\sum} \underset{\substack{ \Delta \mathbf{g}_m \in \Omega_{\Delta \mathbf{g}_m}^{(i)} }}{\max} \hspace{-3mm} \tau_i \widehat{R}_{m,k}\ui \leq R_{\text{leak},m,k},\hspace{1mm}\forall m,\hspace{1mm}\forall k,
\end{align}
where $P_m$ denotes the power consumption of the $m$-th BS and is given by
\begin{align}
P_m = \underset{i \in \mathcal{I}}{\sum} \tau_i \Bigg( \underset{k \in \mathcal{K}}{\sum} \left\| \mathbf{w}_{m,k}\ui \right\|^2 + \text{Tr}(\mathbf{R}_m\ui) \Bigg).
\end{align}
Constraint C1 ensures that the transmit power of the $m$-th BS does not exceed the transmit power budget $P_{\text{max}}$. Constraint C2 guarantees that the achievable rate of the $(m,k)$-th user is larger than or equal to the quality-of-service (QoS) requirement $R_{\text{info},m,k}$ and constraint C3 limits the information leakage rate of $(m,k)$-th user to be smaller than the security requirement $R_{\text{leak},m,k}$.
\par
The non-convexity of problem \eqref{main_prob} stems from the fractional structure in constraints C2 and C3. Furthermore, the optimization variables $\mathbf{w}_{m,k}\uu$ and $\mathbf{R}_m\uu$ are non-trivially involved in the uncertainty set $\Omega_{\Delta \mathbf{g}_m}\ii$ in constraint C3. In addition, the inter-BS interference in the achievable rate and information leakage rate makes it difficult to decompose the problem for decentralized design.
\begin{remark}
Different from the existing sensing-aided secure
communication methods, where the sensing performance requirement is imposed explicitly and separately from the security requirement, limiting the system from fully exploiting the sensing capability to enhance the
overall system performance.
In our proposed adaptive sensing performance design, the sensing performance, i.e., $\mathbf{Q}\ii$, is implicitly formulated in the security requirement specified in constraint C3, enabling the system to adjustably determine an appropriate sensing performance via optimization.
\end{remark}
\section{Centralized Design}
In this section, we consider the centralized design. For that purpose, we first derive the CRB as the sensing performance metric. Then, the beamforming matrices for the communication users and eavesdropper at all BSs are jointly optimized by the central controller.
\label{section_central}
\subsection{Centralized Sensing}
\label{centralized_sensing}
For centralized sensing, location estimation is performed based on the received echoes from all BSs.
The received sensing signal at the $m$-th BS is given by\footnote{BSs can obtain prior knowledge of the eavesdropper's state information from the previous frame using a Kalman filter \cite{kim2018introduction}. Leveraging this information, the BSs are able to identify echoes originating from the eavesdropper using matched filtering and data association based on parameters such as angle of arrival, delay, and other relevant features \cite{210094, 9171304}.}
\begin{align}
\mathbf{y}_{\text{s},m}[l] = \underset{m' \in \mathcal{M}}{\sum} \mathbf{G}_{m',m} \mathbf{x}_{m'}\uu[l] + \mathbf{n}_{\text{s}}[l],
\end{align}
where $\mathbf{G}_{m',m} \overset{\triangle}{=} \alpha_{m',m} \mathbf{a}(\theta_m) \mathbf{a}^H(\theta_{m'})$ is the target response matrix of the link from the $m'$-th BS to the eavesdropper and then to the $m$-th BS, with $\alpha_{m',m}$ denoting the complex channel gain including the radar cross section (RCS). $\mathbf{n}_{\text{s}} \sim \mathcal{CN}(\mathbf{0}, \sigma_s^2 \mathbf{I}_N)$ denotes the AWGN. We note that sensing is performed in the first stage. Hence, $\mathbf{x}_{m'}\uu[l]$ is involved in location estimations. 
\par
By collecting $\mathbf{y}_{\text{s},m}[l]$ at the central controller as
\begin{align}
\mathbf{y}_{\text{s}}^c[l] = [\mathbf{y}_{\text{s},1}^T[l], \dots, \mathbf{y}_{\text{s},M}^T[l]]^T \in \mathbb{C}^{MN \times 1},
\end{align}
we have
\begin{align}
\mathbf{y}_{\text{s}}^c[l] = 
\begin{bmatrix}
\underset{m' \in \mathcal{M}}{\sum} \mathbf{G}_{m',1} \mathbf{x}_{m'}^{(1)}[l] \\
\vdots \\
\underset{m' \in \mathcal{M}}{\sum} \mathbf{G}_{m',M} \mathbf{x}_{m'}^{(1)}[l] \\
\end{bmatrix} + 
\mathbf{n}_{\text{s}}^c[l], \label{y_c}
\end{align}
where $\mathbf{n}_{\text{s}}^c[l] = \mathbf{1}_M \otimes \mathbf{n}_{\text{s}} \in \mathbb{C}^{MN \times 1}$. 
\eqref{y_c} can be compactly reformulated as
\begin{align}
\mathbf{y}_{\text{s}}^c[l] = \underset{m \in \mathcal{M}}{\sum} \mathbf{G}_{m} \widetilde{\mathbf{x}}_m\uu[l] + \mathbf{n}_{\text{s}}^c[l],
\end{align}
where
\begin{align}
& \mathbf{G}_m  = 
\begin{bmatrix}
\mathbf{G}_{m,1} & \mathbf{0} & \cdots & \mathbf{0} \\
\mathbf{0} & \mathbf{G}_{m,2} & \cdots & \mathbf{0} \\
\vdots & \vdots & \ddots & \vdots \\
\mathbf{0} & \mathbf{0} & \cdots & \mathbf{G}_{m,M}
\end{bmatrix} \in \mathbb{C}^{MN \times MN}, \\
& \widetilde{\mathbf{x}}_m\uu[l] = \mathbf{1}_{M} \otimes \mathbf{x}_m\uu[l] \in \mathbb{C}^{MN \times 1}.
\end{align}
The sensing estimation is performed based on the received signals collected over $L$ time slots in the first stage, given by
\begin{align}
\mathbf{Y}_{\text{s}}^c = [\mathbf{y}_{\text{s}}^c[1], \dots, \mathbf{y}_{\text{s}}^c[L]] \in \mathbb{C}^{MN \times L}.
\end{align}
Here, $L$ is predetermined by the system configuration.\footnote{In this paper, we assume that $L$ is fixed. However, in some studies focusing on time-allocation design, $L$ can be optimized \cite{10349846}.} 
The $(i,j)$-th element of Fisher information
matrix (FIM) for estimating the eavesdropper's location $\mathbf{p}$ based on $\mathbf{Y}_{\text{s}}^c$ is given by
\cite{1703855}
\begin{align}
[\mathbf{F}_{\text{central}}]_{i,j} = \frac{2 L}{\sigma_s^2} \Re \left\{ \mathrm{Tr}\left( \underset{m \in \mathcal{M}}{\sum} \dot{\mathbf{G}}^{(j)}_{m} \widetilde{\mathbf{S}}_m\uu \big(\dot{\mathbf{G}}^{(i)}_{m}\big)^H \right) \right\},
\end{align}
where
\begin{align}
\widetilde{\mathbf{S}}_m\uu = (\mathbf{1}_M \mathbf{1}_M^H) \otimes  \mathbf{S}_m\uu.
\end{align}
Matrix $\dot{\mathbf{G}}^{(i)}_{m} \in \mathbb{C}^{MN \times MN}$ denotes the derivative of $\mathbf{G}_{m}$ with respect to $p_i$.
The CRB for estimating $\mathbf{p}$ is given by
\begin{align} \label{Q_central}
\mathbf{Q}_{\text{central}}= \mathbf{F}^{-1}_{\text{central}}.
\end{align} 
For centralized design, $\mathbf{Q}\ii$ in $\Omega_{\Delta \mathbf{g}_m}\ii$ refers to $\mathbf{Q}_{\text{central}}$.
\subsection{Centralized Optimization for Solving Problem \eqref{main_prob}}
By substituting the derived CRB $\mathbf{Q}_{\text{central}}$ into the uncertainty set in constraint C3, we develop the centralized algorithm for solving problem \eqref{main_prob}. 
To begin with, we define $\mathbf{H}_{m,m',k} = \mathbf{h}_{m,m',k} \mathbf{h}_{m,m',k}^H$, $\mathbf{W}_{m,k}^{(i)} = \mathbf{w}_{m,k}^{(i)}(\mathbf{w}_{m,k}^{(i)})^H$ and introduce the rank-one constraint $\mbox{C4:} \hspace{1mm} \mathrm{Rank}(\mathbf{W}_{m,k}) \leq 1, \forall m, \forall k$. 
\par
To handle constraint C2, we first reformulate the achievable rate $R_{m,k}^{(i)}$ as a difference of convex functions, given by
\begin{align}
R_{m,k}^{(i)} = X_{m,k}^{(i)} - Y_{m,k}^{(i)},
\end{align}
where
\begin{align}
X_{m,k}^{(i)} & = \log_2\Big( \mathrm{Tr}\big( \mathbf{H}_{m,m,k} \mathbf{W}_{m,k}\ui \big) + Z_{m,k}\ui \Big), \\
Y_{m,k}^{(i)} & = \log_2 Z_{m,k}\ui.
\end{align}
Here, $Z_{m,k}\ui$ is given by \eqref{Z}, shown at the top of the next page. Next, we adopt successive convex approximation (SCA) to construct an upper bound on the term $Y_{m,k}^{(i)}$ based on first-order Taylor approximation, which provides a sufficient approximation and is given by
\begin{figure*}[t]
\setcounter{equation}{32}
\begin{align} \label{Z}
Z^{(i)}_{m,k} = & \underset{k' \in \mathcal{K} \setminus \{k\} }{\sum} \hspace{-3mm} \mathrm{Tr} \left( \mathbf{H}_{m,m,k} \mathbf{W}_{m,k'}^{(i)} \right) + \hspace{-5mm} \underset{m' \in \mathcal{M} \setminus \{m\}}{\sum} \underset{k' \in \mathcal{K} }{\sum} \mathrm{Tr} \left( \mathbf{H}_{m',m,k} \mathbf{W}_{m',k'}^{(i)} \right) + \hspace{-3mm} \underset{m'\in \mathcal{M}}{\sum} \mathbf{h}_{m',m,k}^H \mathbf{R}_{m'}^{(i)} \mathbf{h}_{m',m,k} + \sigma_{m,k}^2.
\end{align}
\hrule
\setcounter{equation}{33}
\end{figure*}
\begin{align}
& Y_{m,k}^{(i)} \leq Y_{m,k}^{(i)}\left(\ul{\mathbf{W}}^{(i)}, \ul{\mathbf{R}}^{(i)}\right) \notag \\
& +\hspace*{-2mm} \underset{\overline{m}\in \mathcal{M}}{\sum} \underset{\overline{k} \in \mathcal{K}}{\sum} \mathrm{Tr}\hspace*{-0.5mm}\left( \nabla^H_{\mathbf{W}_{\overline{m},\overline{k}}^{(i)}} Y_{m,k}^{(i)}\left(\ul{\mathbf{W}}^{(i)}, \ul{\mathbf{R}}^{(i)}\right) \hspace*{-1mm}\left( \mathbf{W}_{\overline{m},\overline{k}}^{(i)} - \underline{\mathbf{W}}_{\overline{m},\overline{k}}^{(i)} \right) \right) \notag \\
& +\hspace*{-2mm} \underset{\overline{m}\in \mathcal{M}}{\sum} \mathrm{Tr}\left( \nabla^H_{\mathbf{R}_{\overline{m}}^{(i)}} Y_{m,k}^{(i)} \left(\ul{\mathbf{W}}^{(i)}, \ul{\mathbf{R}}^{(i)}\right) \left( \mathbf{R}_{\overline{m}}^{(i)} - \underline{\mathbf{R}}_{\overline{m}}^{(i)} \right) \right) \notag \\
& \overset{\triangle}{=} \overline{Y}_{m,k}^{(i)},
\end{align}
where we define $\mathbf{W}\ui \overset{\triangle}{=} \left\{ \mathbf{W}_{m,k}\ui \right\}_{\forall m, \forall k}$ and $\mathbf{R}\ui \overset{\triangle}{=} \left\{ \mathbf{R}_m\ui \right\}_{\forall m}$, and denote $\underline{\mathbf{W}}\ui$ and $\underline{\mathbf{R}}\ui$ as the values of $\mathbf{W}\ui$ and $\mathbf{R}\ui$ obtained in the previous iteration, respectively.
$\nabla^H_{\mathbf{W}_{\overline{m},\overline{k}}^{(i)}} Y_{m,k}^{(i)}\left(\ul{\mathbf{W}}^{(i)}, \ul{\mathbf{R}}^{(i)}\right)$ and
$\nabla^H_{\mathbf{R}_{\overline{m}}^{(i)}} Y_{m,k}^{(i)} \left(\ul{\mathbf{W}}^{(i)}, \ul{\mathbf{R}}^{(i)}\right)$
denote the gradients of $Y_{m,k}^{(i)}$ with respect to $\mathbf{W}_{\overline{m},\overline{k}}^{(i)}$ and $\mathbf{R}_{\overline{m}}^{(i)}$, respectively, evaluated by $\ul{\mathbf{W}}^{(i)}$ and $\ul{\mathbf{R}}^{(i)}$, given by
\begin{align}
\nabla^H_{\mathbf{W}_{\overline{m},\overline{k}}^{(i)}} Y_{m,k}^{(i)} & = 
\begin{cases}
0 & \hspace{-3mm}: \overline{m} = m, \overline{k} = k, \\
\frac{1}{\ln 2} \frac{\mathbf{H}_{m,m,k}}{Z_{m,k}\ui} & \hspace{-3mm}: \overline{m} = m, \overline{k} \neq k, \\
\frac{1}{\ln 2} \frac{\mathbf{H}_{\overline{m},m,k}}{Z_{m,k}\ui} & \hspace{-3mm}: \text{otherwise},
\end{cases} \\
\nabla^H_{\mathbf{R}_{\overline{m}}^{(i)}} Y_{m,k}^{(i)} & = \frac{1}{\ln 2} \frac{\mathbf{H}_{\overline{m},m,k}}{Z_{m,k}\ui}.
\end{align}
Hence, the lower bound of $R_{\text{info},m,k}^{(i)}$ is given by
\begin{align}
R_{\text{info},m,k}^{(i)} \geq X_{m,k}^{(i)} - \overline{Y}_{m,k}^{(i)}.
\end{align}
As a result, constraint C2 is transformed into
\begin{align}
\overline{\mbox{C2}} \mbox{:} \hspace{1mm} \underset{i \in \mathcal{I}}{\sum} \tau_i \big( X_{m,k}^{(i)} - \overline{Y}_{m,k}^{(i)} \big) \geq R_{\text{info},m,k},\hspace{1mm}\forall m,\hspace{1mm}\forall k.
\end{align}
\par
Due to the existence of the CSI uncertainty, the transformation devised for constraint C2 does not apply to constraint C3. To handle this, we introduce auxiliary variables $\xi_{m,k}^{(i)}$ and the following constraint
\begin{align} \label{C5}
\mbox{C5:} \hspace{1mm} \xi_{m,k}^{(i)} \geq \underset{\substack{ \Delta \mathbf{g}_m \in \Omega_{\Delta \mathbf{g}_m}^{(i)} }}{\max} \frac{ \mathbf{g}_m^H \mathbf{W}_{m,k}^{(i)} \mathbf{g}_m }{ \underset{m \in \mathcal{M}}{\sum} \mathbf{g}_m^H \mathbf{R}_m^{(i)} \mathbf{g}_m + \sigma_{\text{e}}^2 },\hspace{1mm}\forall m,\hspace{1mm}\forall k.
\end{align}
Then, constraint C3 is formulated as 
\begin{align}
\mbox{C3} \mbox{:} \hspace{1mm} \underset{i \in \mathcal{I}}{\sum} \tau_i \log_2 \big(1 + \xi_{m,k}^{(i)}\big) \leq R_{\text{leak},m,k},\hspace{1mm}\forall m,\hspace{1mm}\forall k.
\end{align}
Here, $\log_2 \big(1 + \xi_{m,k}^{(i)}\big)$ is concave. Hence, C3 is essentially a sublevel set of a difference of convex functions, which is non-convex \cite{boyd2004convex}. We employ SCA to handle this issue by constructing an upper bound of $\log_2 \big(1 + \xi_{m,k}^{(i)}\big)$ given by
\begin{align}
\log_2 \big(1 + \xi_{m,k}^{(i)}\big) & \leq  \log_2 \big(1 + \ul{\xi}_{m,k}^{(i)}\big) \notag \\
& + \frac{1}{\ln(2) (1+\underline{\xi}_{m,k}^{(i)})} ( \xi_{m,k}^{(i)} - \underline{\xi}_{m,k}^{(i)} ),
\end{align}
where $\underline{\xi}_{m,k}^{(i)}$ denotes the solution of $\xi_{m,k}^{(i)}$ obtained in the previous iteration. Constraint C3 is then transformed into
\begin{align} \label{c_leak}
\overline{\mbox{C3}} \mbox{:} \hspace{1mm} & \underset{i \in \mathcal{I}}{\sum} \tau_i \Big(\hspace*{-0.5mm} \log_2 \big(1 + \ul{\xi}_{m,k}^{(i)}\big) + \frac{1}{\ln(2) (1\hspace{-0.5mm}+\hspace{-0.5mm}\underline{\xi}_{m,k}^{(i)})} ( \xi_{m,k}^{(i)} - \underline{\xi}_{m,k}^{(i)} ) \Big) \notag \\
& \leq R_{\text{leak},m,k},\hspace{1mm}\forall m, \hspace{1mm}\forall k.
\end{align}
\par
Next, we handle constraint C5 by first eliminating the fractional formulation, given by
\begin{align} \label{C4_2}
\mbox{C5:} \hspace{1mm} & \mathbf{g}_m^H \mathbf{W}_{m,k}^{(i)} \mathbf{g}_m  - \xi_{m,k}^{(i)} \underset{m \in \mathcal{M}}{\sum} \mathbf{g}_m^H \mathbf{R}_m^{(i)} \mathbf{g}_m - \xi_{m,k}^{(i)} \sigma^2_{\text{e}} \leq 0, \notag \\ 
& \forall m,\hspace{1mm}\forall k,\hspace{1mm}\forall \Delta \mathbf{g}_m \in \Omega_{\Delta \mathbf{g}_m}^{(i)}.
\end{align}
Substituting \eqref{channel_g} into \eqref{C4_2}, constraint C5 can be compactly reformulated as
\begin{align} \label{C4_compact}
& \mbox{C5:} \hspace{1mm} \Delta \mathbf{g}^H \mathbf{M}^{(i)}_{m,k} \Delta \mathbf{g} + 2 \Re \left\{ \overline{\mathbf{g}}^H \mathbf{M}^{(i)}_{m,k} \Delta \mathbf{g} \right\} + \overline{\mathbf{g}}^H \mathbf{M}^{(i)}_{m,k} \overline{\mathbf{g}}  \notag \\
& - \xi_{m,k}^{(i)} \sigma^2_{\text{e}} \leq 0, \hspace{1mm}\forall m,\hspace{1mm}\forall k,\hspace{1mm}\forall \Delta \mathbf{g} \in \Omega_{\Delta \mathbf{g}}^{(i)},
\end{align}
where we define
\begin{align}
\overline{\mathbf{g}} & \overset{\triangle}{=} \left[ \overline{\mathbf{g}}_1^T, \dots, \overline{\mathbf{g}}_M^T \right]^T \in \mathbb{C}^{MN \times 1}, \notag \\
\Delta \mathbf{g} & \overset{\triangle}{=} \left[ \Delta \mathbf{g}_1^T, \dots,  \Delta \mathbf{g}_M^T \right]^T \in \mathbb{C}^{MN \times 1}, \notag \\
\mathbf{R}^{(i)} & \overset{\triangle}{=} 
\begin{bmatrix}
\mathbf{R}_1^{(i)} & \mathbf{0} & \cdots & \mathbf{0} \\
\mathbf{0} & \mathbf{R}_2^{(i)} & \cdots & \mathbf{0} \\
\vdots & \vdots & \ddots & \vdots \\
\mathbf{0} & \mathbf{0} & \cdots & \mathbf{R}_M^{(i)}
\end{bmatrix} \in \mathbb{C}^{MN \times MN},
\end{align}
and $\mathbf{M}^{(i)}_{m,k} \overset{\triangle}{=} \overline{\mathbf{W}}_{m,k}^{(i)} - \xi_{m,k}^{(i)} \mathbf{R}^{(i)}$. Here,
$\overline{\mathbf{W}}_{m,k}^{(i)} \in \mathbb{C}^{MN \times MN}$ is the block diagonal matrix with the $m$-th diagonal block being $\mathbf{W}^{(i)}_{m,k}$, given by
\begin{align}
\left[ \overline{\mathbf{W}}_{m,k}^{(i)} \right]_{(m-1)N+1:mN, (m-1)N+1:mN} = \mathbf{W}^{(i)}_{m,k},
\end{align}
while the other elements of $\overline{\mathbf{W}}_{m,k}^{(i)}$ are zero. $\Omega_{\Delta \mathbf{g}}\ui$ is the uncertainty set of $\Delta \mathbf{g}$ given by
\begin{align}
\Omega_{\Delta \mathbf{g}}\ui \overset{\triangle}{=} \left\{ \Delta \mathbf{g} : \| \Delta \mathbf{g} \| \leq \beta_{\mathbf{g}}\ui \right\},
\end{align}
where $\beta_{\mathbf{g}}\ui = \sqrt{\underset{m \in \mathcal{M}}{\sum} ( \beta_{\mathbf{g}_m}\ui )^2} $.
\par
Next, we transform constraint C5 into a tractable form by exploiting the following lemma.
\begin{lemma}
\textit{(S-Procedure Lemma \cite{boyd2004convex})} Define two functions $f_i(\mathbf{t}): \mathbb{C}^{N\times 1}\to \mathbb{R}$, $i\in \left \{ 1,2 \right \}$ as
\begin{equation}
f_i(\mathbf{t})= \mathbf{t}^H\mathbf{A}_i\mathbf{t}+2\Re\left \{\mathbf{b}^H_i\mathbf{t}  \right \} + c_i,
\end{equation}
where $\mathbf{A}_i\in \mathbb{H}^N$, $\mathbf{b}_i\in \mathbb{C}^{N\times 1}$, and $\mathrm{c}_i\in \mathbb{R}$. Then, the implication $f_1(\mathbf{t})\leq0 \Rightarrow f_2(\mathbf{t})\leq0$ holds if and only if there exists a variable $\kappa \geq 0$ such that
\begin{equation}
\kappa 
\begin{bmatrix}
\mathbf{A}_1 &  \mathbf{b}_1\\
\mathbf{b}_1^H &  \mathit{c}_1
\end{bmatrix}-\begin{bmatrix}
\mathbf{A}_2 &  \mathbf{b}_2\\
\mathbf{b}_2^H &  \mathit{c}_2
\end{bmatrix}\succeq \mathbf{0}.
\end{equation}
\end{lemma}
According to Lemma 1, the following implication
\begin{align}
\Delta \mathbf{g}^H \Delta \mathbf{g} \leq ( \beta_{\mathbf{g}_m}\ui )^2 \Rightarrow \mbox{C5}
\end{align}
holds if and only if there exists $\eta_{m,k}\ui \geq 0$ satisfying
\begin{align}
\overline{\mbox{C5}} \mbox{:} \hspace{1mm} & \begin{bmatrix}
\eta_{m,k}^{(i)} \mathbf{I}_{MN} & \mathbf{0} \\
\mathbf{0} & -\eta_{m,k}^{(i)} ( \beta_{\mathbf{g}}\ui )^2 + \xi_{m,k}^{(i)} \sigma^2_{\text{e}} 
\end{bmatrix}
\notag \\
& -
\mathbf{B}^H
\mathbf{M}^{(i)}_{m,k}
\mathbf{B} \succeq \mathbf{0}, \forall m, \forall k,  
\end{align}
where $\mathbf{B} \overset{\triangle}{=} \begin{bmatrix}
\mathbf{I}_{MN} & \overline{\mathbf{g}}
\end{bmatrix} $.
\begin{remark}
We note that $\beta_{\mathbf{g}}\uu$ is a constant known to the system, while $\beta_{\mathbf{g}}\ii$ denotes the CSI error of $\mathbf{g}$ depending on the sensing performance obtained in the first stage, and is a function of the optimization variables $\mathbf{W}_{m,k}\uu$ and $\mathbf{R}_m\uu$. In this way, $\mathbf{W}_{m,k}\uu$ and $\mathbf{R}_m\uu$ are optimized to ensure suitable sensing performance for achieving the best system performance.
\end{remark}
As discussed in Remark 3, we separate constraint $\overline{\mbox{C5}}$ by index of the stage $i$, and denote constraint $\overline{\mbox{C5}}$ with $i=1$ and $i=2$ as $\overline{\mbox{C5a}}$ and $\overline{\mbox{C5b}}$, respectively. It can be easily verified that constraint $\overline{\mbox{C5a}}$ is convex. Next, we handle constraint $\overline{\mbox{C5b}}$, where the non-convexity comes from the fact that the optimization variables $\mathbf{W}_{m,k}\uu$ and $\mathbf{R}_m\uu$ are implicitly contained in $\beta_{\mathbf{g}}\ii$ and are coupled with variable $\eta_{m,k}\ii$. First, we introduce  auxiliary variables $\delta_0 \geq 0$, $\delta_m \geq 0, \forall m$, and the following constraints
\begin{align}
\mbox{C6:}\hspace{1mm} \delta_0 \geq \underset{m \in \mathcal{M}}{\sum} \delta_m^2,\hspace{4mm} 
\mbox{C7:}\hspace{1mm} \delta_m \geq \beta_{\mathbf{g}_m}^{(2)},\hspace{1mm}\forall m. \label{C7}
\end{align}
Constraint $\overline{\mbox{C5b}}$ is now rewritten as
\begin{align}
\myoverline{\mbox{C5b}} \mbox{:}\hspace{1mm} & \begin{bmatrix}
\eta_{m,k}^{(2)} \mathbf{I}_{MN} & \mathbf{0} \\
\mathbf{0} & -\eta_{m,k}^{(2)} \delta_0 + \xi_{m,k}^{(2)} \sigma^2_{\text{e}} 
\end{bmatrix} \notag \\
&\hspace{4mm}  - \mathbf{B}^H
\mathbf{M}^{(2)}_{m,k}
\mathbf{B}\hspace{1mm} \succeq\hspace{1mm} \mathbf{0}, \hspace{1mm} \forall m,\hspace{1mm}\forall k.
\end{align}
Substituting \eqref{beta_gm} and \eqref{Q_central} into \eqref{C7}, constraint C7 is reformulated as
\begin{align}
\overline{\mbox{C7}} \mbox{:} \hspace{1mm} \mathrm{Tr}(\mathbf{F}_{\text{central}}^{-1}) - a_m^2 ( \delta_m - b_m )^2 \leq 0, \forall m,
\end{align}
where
\begin{align}
a_m & = \frac{\overline{d}_m }{\alpha_m \sqrt{\frac{\kappa}{1+\kappa}} \sqrt{\frac{N(N-1)(2N-1)}{6}} \frac{3 \pi \sin \overline{\theta}_m }{\overline{d}_m}}, \label{a_m} \\
b_m & = \alpha_m \beta_{\text{NLoS},m} \sqrt{\frac{1}{1+\kappa}}. \label{b_m}
\end{align}
Constraint $\overline{\mbox{C7}}$ is in the canonical form of a difference of convex functions. To tackle this issue, we construct a lower bound of $a_m^2 ( \delta_m - b_m )^2$ based on the first-order Taylor approximation, given by
\begin{align}
& a_m^2 ( \delta_m - b_m )^2 \notag \\
&\hspace{2mm} \geq a_m^2 ( \ul{\delta}_m - b_m )^2 + 2 a_m^2 ( \ul{\delta}_m - b_m )(\delta_m - \ul{\delta}_m),
\end{align}
where $\ul{\delta}_m$ denotes the solution to $\delta_m$ obtained in the previous iteration. Constraint $\overline{\mbox{C7}}$ is transformed into
\begin{align}
\overline{\mbox{C7}} \mbox{:} \hspace{1mm} & \mathrm{Tr}(\mathbf{F}_{\text{central}}^{-1}) - 2 a_m^2 ( \ul{\delta}_m - b_m )(\delta_m - \ul{\delta}_m) \notag \\
&\hspace{2mm}  - a_m^2 ( \ul{\delta}_m - b_m )^2 \leq 0, \hspace{1mm} \forall m.
\end{align}
By employing the semidefinite relaxation (SDR) technique, the rank-one constraint is relaxed \cite{9183907}. The relaxed optimization problem is given by
\begin{align} \label{bcd}
\underset{\substack{ \mathbf{W}_{m,k}^{(i)}, \mathbf{R}_m^{(i)}, \\ \xi_{m,k}^{(i)}, \eta_{m,k}^{(i)} \\ \delta_0, \delta_m }}{\mino} & \hspace*{3mm} \underset{m \in \mathcal{M}}{\sum} P_m \notag\\
\subto & \hspace*{3mm} \mbox{C1}, \overline{\mbox{C2}}, \overline{\mbox{C3}}, \overline{\mbox{C5a}}, \myoverline{\mbox{C5b}}, \mbox{C6}, \overline{\mbox{C7}},
\end{align}
which is still non-convex due to the variable coupling issue. To handle this, we utilize BCD to decouple the variables into two blocks $\left\{ \mathbf{W}_{m,k}^{(i)}, \mathbf{R}_m^{(i)}, \delta_0, \delta_m \right\}$ and $\left\{ \xi_{m,k}^{(i)}, \eta_{m,k}^{(i)} \right\}$ \cite{xu2025resolution}. 
The optimization problem with the block $\left\{ \mathbf{W}_{m,k}^{(i)}, \mathbf{R}_m^{(i)}, \delta_0, \delta_m \right\}$ is given by
\begin{align} \label{bcd_1}
\underset{\substack{ \mathbf{W}_{m,k}^{(i)}, \mathbf{R}_m^{(i)}, \\ \delta_0, \delta_m }}{\mino} & \hspace*{3mm} \underset{m \in \mathcal{M}}{\sum} P_m \notag\\
\subto & \hspace*{3mm} \mbox{C1}, \overline{\mbox{C2}}, \overline{\mbox{C5a}}, \myoverline{\mbox{C5b}}, \mbox{C6}, \overline{\mbox{C7}},
\end{align}
which is convex and can be efficiently solved by standard convex
optimization solvers such as CVX \cite{grant2014cvx}.
\par
The optimization problem with respect to block $\left\{ \xi_{m,k}^{(i)}, \eta_{m,k}^{(i)} \right\}$ is a feasibility check problem given by
\begin{align} \label{feasibilty_check}
\underset{\substack{ }}{\mathrm{Find}} & \hspace*{3mm} \xi_{m,k}^{(i)}, \eta_{m,k}^{(i)} \notag\\
\subto & \hspace*{3mm} \overline{\mbox{C3}}, \overline{\mbox{C5a}}, \myoverline{\mbox{C5b}}.
\end{align}
To accelerate the convergence and to provide more DoFs for optimizing block $\left\{ \mathbf{W}_{m,k}^{(i)}, \mathbf{R}_m^{(i)}, \delta_0, \delta_m \right\}$, problem \eqref{feasibilty_check} can be replaced by solving the following information leakage rate minimization problem for updating $\left\{ \xi_{m,k}^{(i)}, \eta_{m,k}^{(i)} \right\}$, given by
\begin{align} \label{bcd_2}
\underset{\substack{ \xi_{m,k}^{(i)}, \eta_{m,k}^{(i)} }}{\mino} & \hspace*{3mm} \underset{i \in \mathcal{I}}{\sum} \tau_i \Big( \log_2 \big(1 + \ul{\xi}_{m,k}^{(i)}\big) + \notag \\
& \hspace*{3mm} \frac{1}{\ln(2) (1+\underline{\xi}_{m,k}^{(i)})} ( \xi_{m,k}^{(i)} - \underline{\xi}_{m,k}^{(i)} ) \Big) \notag \\
\subto & \hspace*{3mm} \overline{\mbox{C5a}}, \myoverline{\mbox{C5b}},
\end{align}
which is a convex optimization problem and can be solved efficiently. 
\par
The centralized design for solving problem \eqref{main_prob} is summarized in $\textbf{Algorithm 1}$, where $f_1^{(n)}$ denotes the objective function value of problem \eqref{bcd_1} at the $n$-th iteration. The convergence and computational complexity of Algorithm 1 are discussed as follows.
\begin{algorithm}[t]
\caption{Centralized Algorithm for Solving Problem \eqref{main_prob}}
\begin{algorithmic}[1]
\small
\STATE Set iteration index $n=1$, error tolerances $\epsilon_1$.
\REPEAT
\STATE Update $\mathbf{W}_{m,k}^{(i)}, \mathbf{R}_m^{(i)}, \delta_0, \delta_m$ by solving problem \eqref{bcd_1}
\STATE Update $\xi_{m,k}^{(i)}, \eta_{m,k}^{(i)}$ by solving problem \eqref{bcd_2}
\STATE Set $n \leftarrow n + 1$
\UNTIL $\frac{\left|f_1^{(n)} - f_1^{(n-1)}\right|}{f_1^{(n-1)}} \leq \epsilon_1$
\end{algorithmic}
\end{algorithm}
\par
By alternatively solving problem \eqref{bcd_1} and problem \eqref{bcd_2}, the objective function value of problem \eqref{bcd} is non-increasing, and the corresponding solution is guaranteed to converge to a stationary point
of optimization problem \eqref{main_prob} \cite{razaviyayn2013unified}. The computational complexity of Algorithm 1 is dominated by solving the SDP problem \eqref{bcd_1} and is given by $\mathcal{O} \big( \log\frac{1}{\varepsilon} (M^4KN^3 + M^4K^2N^2 +M^3K^3) \big)$, where $\varepsilon$ is the convergence tolerance of the interior point
method for solving the SDP problem \cite[Theorem 3.12]{bomze2010interior}.

\section{Decentralized Design}
The centralized design requires collecting the CSI and sensing echoes at all BSs and optimizing the optimization variables at all BSs together, leading to a high information exchange overhead and computational complexity. To alleviate this issue, this section presents a decentralized design for solving problem \eqref{main_prob}, where each BS individually designs its transmit beamforming only based on the local CSI and sensing echoes. For that purpose, we first derive the CRB under the decentralized sensing scheme. We then divide the optimization problem into $M$ subproblems, each of which corresponds to the information beamforming and radar covariance matrix of one BS. As a result, the overhead and the computational complexity are vastly reduced.

\subsection{Decentralized Sensing}
\label{decentralized_sensing}
For decentralized sensing, each BS estimates the location of the eavesdropper based on its local sensing echoes. The final location estimation result is obtained by fusing the estimation results of all BSs. For the $m$-th BS, by collecting its received echoes $\mathbf{y}_{\text{s}, m}[l]$ during $L$ time slots in the first stage, we obtain
\begin{align}
\mathbf{Y}_{\text{s},m} = \underset{m' \in \mathcal{M}}{\sum} \mathbf{G}_{m',m} \mathbf{X}_{m'}\uu + \mathbf{N}_{\text{s}}, \label{sensing_signal}
\end{align} 
where 
\begin{align}
\mathbf{Y}_{\text{s},m} = [\mathbf{y}_{\text{s}, m}[1], \dots, \mathbf{y}_{\text{s}, m}[L]] \in \mathbb{C}^{N \times L}, \\
\mathbf{X}_m\uu = [\mathbf{x}_{m}\uu[1], \dots, \mathbf{x}_{m}\uu[L]] \in \mathbb{C}^{N \times L}.
\end{align}
By vectorizing \eqref{sensing_signal}, we have
\begin{align}
\widetilde{\mathbf{y}}_{\text{s},m} = \mathrm{vec}(\mathbf{Y}_{\text{s},m}) = \mathbf{u}_m + \widetilde{\mathbf{n}}_{\text{s}}, \label{vec_sensing}
\end{align}
where $\mathbf{u}_m = \mathrm{vec}\left(\underset{m' \in \mathcal{M}}{\sum} \mathbf{G}_{m',m} \mathbf{X}_{m'}\uu \right)$, $\widetilde{\mathbf{n}}_{\text{s}} = \mathrm{vec}(\mathbf{N}_{\text{s}})$ and $\widetilde{\mathbf{n}}_{\text{s}} \sim \mathcal{CN}(\mathbf{0}, \sigma_{\text{s}}^2 \mathbf{I}_{NL})$.
\par
The $(i,j)$-th element of FIM for estimating $\mathbf{p}$ at the $m$-th BS is calculated by
\begin{align}
[\mathbf{F}_m]_{i,j} = & \frac{2 L}{\sigma_s^2} \Re \left\{ \mathrm{Tr}\left( \underset{m' \in \mathcal{M}}{\sum} \dot{\mathbf{G}}^{(j)}_{m',m} \mathbf{S}_{m'} \big(\dot{\mathbf{G}}^{(i)}_{m',m}\big)^H \right) \right\}, \notag \\
& \hspace{3mm} i,j \in \{1,2\},
\end{align}
where $\dot{\mathbf{G}}^{(i)}_{m',m} \in \mathbb{C}^{N \times N}$ denotes the derivative of $\mathbf{G}_{m',m}$ with respect to $p_i$. The corresponding CRB matrix $\mathbf{Q}_m \in \mathbb{R}^{2 \times 2}$ is given by
\begin{align}
\mathbf{Q}_m = \mathbf{F}_m^{-1}.
\end{align}
\par
Denote $\hat{\mathbf{p}}_m$ as the estimated location of the eavesdropper at the $m$-th BS. The joint estimation result can be obtained by fusing $\hat{\mathbf{p}}_m$ and $\mathbf{Q}_m$ collected from all BSs at the central controller.
\begin{remark}
In decentralized sensing, each BS only needs to transmit the estimation results, i.e., $\hat{\mathbf{p}}_m$ and $\mathbf{Q}_m$, resulting in a much lower information exchange overhead compared to centralized sensing, where the original echoes accumulated in $L$ times slots need to be transmitted.
\end{remark}
We adopt the linear estimator for location estimation based on $\hat{\mathbf{p}}_m$ and $\mathbf{Q}_m$, given by \cite{pei2019elementary}
\begin{align} \label{p_fuse}
\hat{\mathbf{p}}_{\text{fused}} = \underset{m \in \mathcal{M}}{\sum} \mathbf{A}_m \hat{\mathbf{p}}_m.
\end{align}
Here, $\mathbf{A}_m$ denotes the fusion weights, which are determined to minimize the mean square error (MSE) of $\hat{\mathbf{p}}_{\text{fused}}$, given by
\begin{align}
\text{MSE}(\hat{\mathbf{p}}_{\text{fused}}) = \mathbb{E}\left[ \underset{m \in \mathcal{M}}{\sum} ( \hat{\mathbf{p}}_m - \mathbf{p} )^T \mathbf{A}_m^T \mathbf{A}_m ( \hat{\mathbf{p}}_m - \mathbf{p} ) \right].
\end{align}
The MSE minimization problem is formulated as
\begin{align}
\hspace*{-3mm}\underset{\substack{ \mathbf{A}_m }}{\mino} \hspace*{3mm} & \text{MSE}(\hat{\mathbf{p}}_{\text{fused}}) \notag\\
\subto\hspace*{3mm} &  \underset{m \in \mathcal{M}}{\sum} \mathbf{A}_m = \mathbf{I}, \label{mse_min}
\end{align}
where the constraint $\underset{m \in \mathcal{M}}{\sum} \mathbf{A}_m = \mathbf{I}$ represents the normalization of the weights. The Lagrangian function of problem \eqref{mse_min} is given by
\begin{align}
\mathcal{L} = & \mathbb{E}\left[ \underset{m \in \mathcal{M}}{\sum} ( \hat{\mathbf{p}}_m - \mathbf{p} )^T \mathbf{A}_m^T \mathbf{A}_m ( \hat{\mathbf{p}}_m - \mathbf{p} ) \right] \notag \\
& + \mathrm{Tr} \left( \bm{\Pi}^T \left( \underset{m \in \mathcal{M}}{\sum} \mathbf{A}_m - \mathbf{I} \right) \right),
\end{align}
where $\bm{\Pi}$ denotes the dual variable corresponding to the constraint.
According to the Karush-Kuhn-Tucker (KKT) condition, we have
\begin{align}
\frac{\partial \mathcal{L}}{\partial \mathbf{A}_m} = \mathbb{E}\left[ 2\mathbf{A}_m \left( \hat{\mathbf{p}}_m - \mathbf{p} \right) \left( \hat{\mathbf{p}}_m - \mathbf{p} \right)^T + \bm{\Pi} \right] = \mathbf{0}.
\end{align}
Hence, $\mathbf{A}_m \mathbf{Q}_m = - \frac{1}{2} \bm{\Pi} $. Together with the constraint $\underset{m \in \mathcal{M}}{\sum} \mathbf{A}_m = \mathbf{I}$, the optimal $\mathbf{A}_m$ is obtained, given by
\begin{align}
\mathbf{A}_m^* = \left( \underset{m \in \mathcal{M}}{\sum} \mathbf{Q}_m^{-1} \right)^{-1} \mathbf{Q}_m^{-1}.
\end{align}
The covariance matrix of $\hat{\mathbf{p}}_{\text{fused}}$, denoted by $\mathbf{Q}_{\text{fused}}$, is given by
\begin{align} \label{Q_fused}
\mathbf{Q}_{\text{fused}} & = \underset{m \in \mathcal{M}}{\sum} \mathbf{A}_m^* \mathbf{Q}_m (\mathbf{A}_m^*)^T =
\left( \underset{m \in \mathcal{M}}{\sum} \mathbf{Q}_m^{-1} \right)^{-1} \notag \\
& = \left( \underset{m \in \mathcal{M}}{\sum} \mathbf{F}_m \right)^{-1}.
\end{align}
For decentralized design, $\mathbf{Q}\ii$ in $\Omega_{\Delta \mathbf{g}_m}\ii$ refers to $\mathbf{Q}_{\text{fused}}$.

\subsection{Decentralized Optimization for Solving Problem \eqref{main_prob}}
By substituting the derived CRB, $\mathbf{Q}_{\text{fused}}$, into the uncertainty set of constraint C3, we develop a decentralized algorithm to solve problem \eqref{main_prob}. The primary challenge arises from the fact that the CSI at each BS is unknown to the others, as each BS retains its own information. Moreover, inter-BS interference further hinders problem decomposition, making decentralized design particularly challenging.
\par
To begin with, we introduce auxiliary variables $e_{m',m,k}$ and $t_{m',m,k}^{(i)}$, referring to the interference caused by the information signals and sensing signals from the $m'$-th BS in the achievable rate of the $(m,k)$-th user, i.e., $R_{m,k}$, given by
\begin{align}
\mbox{C8:} & \hspace{1mm} e_{m',m,k}\ui \geq \hspace{-2mm} \underset{k' \in \mathcal{K} }{\sum} \hspace{-1mm} \mathrm{Tr}\left( \mathbf{H}_{m',m,k} \mathbf{W}_{m',k'}^{(i)} \right), \forall (m,m') \in \overline{\mathcal{M}},\hspace{1mm}\forall k, \label{c_e} \\
\mbox{C9:} & \hspace{1mm} t_{m',m,k}^{(i)} \geq \mathbf{h}_{m',m,k}^H \mathbf{R}_{m'}^{(i)} \mathbf{h}_{m',m,k}, \forall (m,m') \in \overline{\mathcal{M}},\hspace{1mm}\forall k, \label{c_t}
\end{align}
where the set $\overline{\mathcal{M}}$ is defined as $\overline{\mathcal{M}} \overset{\triangle}{=} \{(m,m'): m \neq m', m \in \mathcal{M}, m' \in \mathcal{M} \}$.
Substituting \eqref{c_e} and \eqref{c_t} into \eqref{achievable_rate}, we obtain the lower bound of $R_{m,k}$ based on the monotonicity of the reciprocal function, denoted by $\widetilde{R}_{m,k}^{(i)}\left( \mathbf{W}^{(i)}_{m,k}, \mathbf{R}^{(i)}_m, \left\{ e_{m',m,k}^{(i)} \right\}_{\forall m' \neq m}, \left\{ t_{m',m,k}^{(i)} \right\}_{\forall m' \neq m} \right)$, given in \eqref{wide_R} shown at the
bottom of the next page. Constraint C2 is replaced by
\setcounter{equation}{78}
\begin{align}
\widetilde{\mbox{C2}} \mbox{:} \hspace{1mm} \underset{i \in \mathcal{I}}{\sum} \tau_i \widetilde{R}_{m,k}\ui \geq R_{\text{info},m,k}, \hspace{1mm}\forall m,\hspace{1mm}\forall k.
\end{align}
\par
Similar to the steps \eqref{C5}-\eqref{c_leak} used to handle constraint C3 in Section \ref{section_central}, we introduce auxiliary variables $\xi_{m,k}\ui$ and constraint C5 given in \eqref{C5}. Constraint C3 is transformed into 
\begin{align} \label{c_leak_2}
\widetilde{\mbox{C3}} \mbox{:} \hspace{1mm} \widetilde{\widehat{R}}_{m,k}\ui \leq R_{\text{leak},m,k},\hspace{1mm}\forall m,\hspace{1mm}\forall k,
\end{align}
where $\widetilde{\widehat{R}}_{m,k}\ui \overset{\triangle}{=} \underset{i \in \mathcal{I}}{\sum} \tau_i \Big( \log_2 \big(1 + \ul{\xi}_{m,k}^{(i)}\big) + \frac{1}{\ln(2) (1+\underline{\xi}_{m,k}^{(i)})} ( \xi_{m,k}^{(i)} - \underline{\xi}_{m,k}^{(i)} ) \Big)$.
Then, we introduce auxiliary variable $u_m^{(i)}$ referring to the interference from the $m$-th BS at the eavesdropper, given by
\begin{align}
\mbox{C10} \mbox{:} \hspace{1mm} & u_{m}^{(i)} = \underset{\substack{ \Delta \mathbf{g}_m \in \Omega_{\Delta \mathbf{g}_m}^{(i)} }}{\min} \mathbf{g}_m^H \mathbf{R}_m^{(i)} \mathbf{g}_m, \hspace{1mm}\forall m.
\end{align}
Constraint C5 is then rewritten as
\begin{align}
\mbox{C5} \mbox{:} \hspace{1mm} & \mathbf{g}_m^H \mathbf{W}_{m,k}^{(i)} \mathbf{g}_m  - \xi_{m,k}^{(i)} \mathbf{g}_m^H \mathbf{R}_m^{(i)} \mathbf{g}_m - \xi_{m,k}^{(i)} \underset{m' \in \mathcal{M} \setminus\{m\}}{\sum} u_{m'}^{(i)} \notag \\
&- \xi_{m,k}^{(i)} \sigma^2_{\text{e}}  \leq 0,\hspace{1mm}\forall m,\hspace{1mm}\forall k,\hspace{1mm}\forall \Delta \mathbf{g}_m \in \Omega_{\Delta \mathbf{g}_m}^{(i)}.
\end{align}
By introducing auxiliary variables $\psi_{m,k}^{(i)} \geq 0$ and $\lambda_m^{(i)} \geq 0$ and according to Lemma 1, constraints C5 and C10 are equivalently transformed into
\begin{align}
\widetilde{\mbox{C5}} \mbox{:} \hspace{1mm} & \bm{\Psi}_{m,k}^{(i)}\left( \mathbf{W}^{(i)}_{m,k}, \mathbf{R}^{(i)}_m, \psi_{m,k}^{(i)}, \xi_{m,k}^{(i)}, \left\{ u_{m'}^{(i)} \right\}_{\forall m' \neq m} \right) \succeq \mathbf{0}, \notag \\
&\hspace{1mm}\forall m,\hspace{1mm}\forall k, \\
\widetilde{\mbox{C10}} \mbox{:} \hspace{1mm} & \bm{\Phi}_m^{(i)} \left( \mathbf{R}^{(i)}_m, \lambda_m^{(i)}, u_m^{(i)} \right)  \succeq \mathbf{0},\hspace{1mm}\forall m,
\end{align}
where $\bm{\Psi}_{m,k}^{(i)}\left( \mathbf{W}^{(i)}_{m,k}, \mathbf{R}^{(i)}_m, \psi_{m,k}^{(i)}, \xi_{m,k}^{(i)}, \left\{ u_{m'}^{(i)} \right\}_{m' \neq m} \right)$ and $\bm{\Phi}_m^{(i)} \left( \mathbf{R}^{(i)}_m, \lambda_m^{(i)}, u_m^{(i)} \right)$ are given in \eqref{Psi_eq} and \eqref{Phi_eq}, respectively, shown at the bottom of the next page. Here, we define $\widetilde{\mathbf{M}}^{(i)}_{m,k} = \mathbf{W}^{(i)}_{m,k} -  \xi^{(i)}_{m,k} \mathbf{R}^{(i)}_m $ and $\mathbf{D}_m = \begin{bmatrix} \mathbf{I}_N & \overline{\mathbf{g}}_m \end{bmatrix}$.
\par
Note that $\beta_{\mathbf{g}_m}\ii$ depends on variables $\mathbf{W}_{m,k}\uu$ and $\mathbf{R}_m\uu$, we denote $\widetilde{\mbox{C5}}$ as $\widetilde{\mbox{C5a}}$ for $i=1$, and as $\widetilde{\mbox{C5b}}$ for $i=2$. Also, we denote $\widetilde{\mbox{C10}}$ as $\widetilde{\mbox{C10a}}$ for $i=1$, and as $\widetilde{\mbox{C10b}}$ for $i=2$. 
Then, we introduce variables $\delta_m$ satisfying
\setcounter{equation}{86}
\begin{align} \label{delta_m}
\mbox{C11} \mbox{:} & \hspace{1mm} \delta_m \geq \beta_{\mathbf{g}_m}^{(2)},\hspace{1mm}\forall m.
\end{align}
By substituting \eqref{delta_m}, $\widetilde{\mbox{C5b}}$ and $\widetilde{\mbox{C10b}}$ are transformed into 
\begin{align}
\mywidetilde{\mbox{C5b}} \mbox{:} \hspace{1mm} & \bm{\Psi}_{m,k}^{(2)}\left( \mathbf{W}^{(2)}_{m,k}, \mathbf{R}^{(2)}_m, \psi_{m,k}^{(2)}, \xi_{m,k}^{(2)}, \left\{ u_{m'}^{(2)} \right\}_{\forall m' \neq m}, \delta_m \right) \notag \\ &\succeq \mathbf{0},  
\forall m,\hspace{1mm}\forall k. \\
\mywidetilde{\mbox{C10b}} \mbox{:} \hspace{1mm} & \bm{\Phi}_m^{(2)} \left( \mathbf{R}^{(2)}_m, \lambda_m^{(2)}, u_m^{(2)}, \delta_m \right)  \succeq \mathbf{0},\hspace{1mm}\forall m,
\end{align}
By substituting \eqref{beta_gm} and \eqref{Q_fused} into \eqref{delta_m}, constraint C11 is recast as
\begin{align}
\widetilde{\mbox{C11}} \mbox{:} \hspace{1mm} \mathrm{Tr}\left( \left( \underset{m \in \mathcal{M}}{\sum} \mathbf{F}_m \right)^{-1} \right) - a_m^2 ( \delta_m - b_m )^2
\leq 0,\hspace{1mm}\forall m,
\end{align}
where $a_m$ and $b_m$ are defined in \eqref{a_m} and \eqref{b_m}, respectively. Next, we apply SCA to handle the non-convex term $- a_m^2 ( \delta_m - b_m )^2$, constraint $\widetilde{\mbox{C11}}$ is transformed into
\begin{align}
\widetilde{\mbox{C11}} \mbox{:} \hspace{1mm} & \mathrm{Tr}\left( \left( \underset{m \in \mathcal{M}}{\sum} \mathbf{F}_m \right)^{-1} \right) - 2 a_m^2 ( \ul{\delta}_m - b_m )(\delta_m - \ul{\delta}_m) \notag \\
& - a_m^2 ( \ul{\delta}_m - b_m )^2 \leq 0, \forall m.
\end{align}
Next, by defining
\begin{align}
\left[ \overline{\mathbf{F}}_m \right]_{i,j} = \frac{2 L}{\sigma_s^2} \Re \left\{ \mathrm{Tr}\left( \underset{m' \in \mathcal{M}}{\sum} \dot{\mathbf{G}}^{(j)}_{m,m'} \mathbf{S}_{m}^{(1)} [\dot{\mathbf{G}}^{(i)}_{m,m'}]^H \right) \right\},
\end{align}
we have
\begin{align}
\underset{m \in \mathcal{M}}{\sum} \mathbf{F}_m = \underset{m \in \mathcal{M}}{\sum} \overline{\mathbf{F}}_m.
\end{align}
Furthermore, by introducing $\mathbf{V}_m \in \mathbb{C}^{2 \times 2}$ and constraint
\begin{align}
\mbox{C12:} \hspace{1mm} \overline{\mathbf{F}}_m \succeq \mathbf{V}_m, \forall m,
\end{align}
constraint $\widetilde{\mbox{C11}}$ is rewritten as
\begin{align}
\mywidetilde{\mbox{C11}} \mbox{:} \hspace{1mm} \Pi_m\left( \mathbf{W}_{m,k}^{(1)}, \mathbf{R}_m^{(1)}, \left\{\mathbf{V}_{m'}\right\}_{\forall m' \neq m}, \delta_m \right) \leq 0, \forall m,
\end{align}
where $\Pi_m \left( \mathbf{W}_{m,k}^{(1)}, \mathbf{R}_m^{(1)}, \left\{\mathbf{V}_{m'}\right\}_{\forall m' \neq m}, \delta_m \right)$ is given by \eqref{Pi}, shown at the bottom of the next page.

\par
With SDR relaxation, problem \eqref{main_prob} is now transformed into
\setcounter{equation}{96}
\begin{align} \label{prob_1}
\underset{\substack{ \mathbf{W}_{m,k}^{(i)}, \mathbf{R}_m^{(i)}, \mathbf{V}_m, \\ \xi_{m,k}\ui, \psi_{m,k}\ui, \lambda_m\ui, \delta_m , \\ e_{m',m,k}\ui, t_{m',m,k}\ui, u_m\ui }}{\maxo} & \hspace*{3mm} \underset{m \in \mathcal{M}}{\sum} P_m \notag\\
\subto & \hspace*{3mm} \mbox{C1}, \widetilde{\mbox{C2}}, \widetilde{\mbox{C3}}, \widetilde{\mbox{C5a}}, \mywidetilde{\mbox{C5b}}, \notag \\
& \hspace*{3mm} \mbox{C8}, \mbox{C9}, \widetilde{\mbox{C10a}}, \mywidetilde{\mbox{C10b}}, \mywidetilde{\mbox{C11}}, \mbox{C12}.
\end{align}
We note that although $\mathbf{H}_{m,m',k}^{(i)}$ $\mathbf{W}_{m,k}^{(i)}$ and $\mathbf{R}_m^{(i)}$ are separated in each constraint indexed by $m$ and $k$, the variables $\left\{ e_{m',m,k}^{(i)} \right\}_{\forall m' \neq m}$, $\left\{ t_{m',m,k}^{(i)}\right\}_{\forall m' \neq m}$ in $\widetilde{R}_{m,k}^{(i)}$, $\left\{ u_{m'}^{(i)} \right\}_{\forall m' \neq m}$ in $\bm{\Psi}_{m,k}\ui$, and $\left\{\mathbf{V}_{m'}\right\}_{\forall m' \neq m}$ in $\Pi_m$ hinder independently decomposing the problem for decentralized optimization. To this end, we construct a set of information to be shared among all BSs by introducing auxiliary variables $\mathbf{e}^{(i)} \in \mathbb{R}^{M(M-1)K \times 1}$, $\mathbf{t}^{(i)} \in \mathbb{R}^{M(M-1)K\times 1}$, $\mathbf{u}^{(i)} \in \mathbb{R}^{M \times 1}$, and $\mathbf{V} \in \mathbb{C}^{2M \times 2}$, which are referred to as global variables, given by
\begin{align}
\mathbf{e}^{(i)} &\hspace*{-0.5mm}\overset{\triangle}{=}\hspace*{-0.5mm} \left[ [e_{1,2,1}^{(i)}, \dots, e_{1,2,K}^{(i)}], \dots, [e_{M,M-1,1}^{(i)}, \dots, e_{M,M-1,K}^{(i)}] \right]^T, \label{g1} \\
\mathbf{t}^{(i)} &\hspace*{-0.5mm} \overset{\triangle}{=}\hspace*{-0.5mm} \left[ [t_{1,2,1}^{(i)}, \dots, t_{1,2,K}^{(i)}], \dots, [t_{M,M-1,1}^{(i)}, \dots, t_{M,M-1,K}^{(i)}] \right]^T, \label{g2}\\
\mathbf{u}^{(i)} & \overset{\triangle}{=} \left[ u_{1}^{(i)}, \dots, u_{M}^{(i)} \right]^T, \hspace{3mm}
\mathbf{V} \overset{\triangle}{=} \left[ \mathbf{V}_1^T, \dots, \mathbf{V}_M^T \right]^T. \label{g3}
\end{align}
We define the set $\mathcal{G} \overset{\triangle}{=} \left\{ \mathbf{e}^{(i)}, \mathbf{t}^{(i)}, \mathbf{u}^{(i)}, \mathbf{V} \right\} $ to represent the global variables.
Furthermore, we define
\begin{align}
& \overline{e}_{m,k}^{(i)} = \underset{m' \in \mathcal{M} \setminus \{m\}}{\sum}  e_{m',m,k}^{(i)}, \hspace{3mm}
\overline{t}_{m,k}^{(i)} = \underset{m'\in \mathcal{M}\setminus \{m\}}{\sum}  t_{m',m,k}^{(i)}, \\
& \overline{u}_m^{(i)} = \underset{m' \in \mathcal{M} \setminus \{m\}}{\sum} u_{m'}^{(i)}, \hspace{10mm}
\overline{\mathbf{V}}_m = \underset{m' \in \mathcal{M} \setminus \{m\} }{\sum} \mathbf{V}_{m'}.
\end{align}
The corresponding variables retained at the $m$-th BS are given by
\begin{align} 
\mathbf{e}_m^{(i)} & \overset{\triangle}{=} \big[ [\overline{e}_{m,1}^{(i)}, \dots, \overline{e}_{m,K}^{(i)}], [e_{m,1,1}^{(i)}, \dots, e_{m,1,K}^{(i)}] , \dots \notag \\
& \hspace{8mm} [e_{m,M,1}^{(i)}, \dots, e_{m,M,K}^{(i)}] \big]^T \in \mathbb{R}^{MK \times 1}, \label{e1} \\
\mathbf{t}_m^{(i)} & \overset{\triangle}{=} \big[ [\overline{t}_{m,1}^{(i)}, \dots, \overline{t}_{m,K}^{(i)}], [t_{m,1,1}^{(i)}, \dots, t_{m,1,K}^{(i)}], \dots, \notag \\
& \hspace{8mm} [t_{m,M,1}^{(i)}, \dots, t_{m,M,K}^{(i)}] \big]^T \in \mathbb{R}^{MK \times 1}, \label{t1} \\
\mathbf{u}_m^{(i)} & \overset{\triangle}{=} \left[ \overline{u}_m^{(i)}, u_m^{(i)} \right]^T \in \mathbb{R}^{2 \times 2}, \label{u1} \\
\widehat{\mathbf{V}}_m & \overset{\triangle}{=} \left[ \overline{\mathbf{V}}_m^T, \mathbf{V}_m^T \right]^T \in \mathbb{C}^{4 \times 2}, \label{v1}
\end{align}
which are referred to as the local variables.
The connection between the global variables and local variables is constructed by the following constraints
\begin{align}
&\mbox{C13a:}  \hspace{1mm} \mathbf{e}_m^{(i)} = \mathbf{E}_m \mathbf{e}^{(i)}, \hspace{1mm} \forall m,   \\
&\mbox{C13b:}  \hspace{1mm} \mathbf{t}_m^{(i)} = \mathbf{T}_m \mathbf{t}^{(i)}, \hspace{1mm} \forall m,  \\
&\mbox{C13c:}  \hspace{1mm} \mathbf{u}_m^{(i)} = \mathbf{U}_m \mathbf{u}^{(i)}, \hspace{1mm} \forall m,   \\
&\mbox{C13d:}  \hspace{1mm} \widehat{\mathbf{V}}_m = \widetilde{\mathbf{V}}_m \mathbf{V}, \hspace{1mm} \forall m,
\end{align}
where the linear mappings $\mathbf{E}_m$, $\mathbf{T}_m$, $\mathbf{U}_m$, and $\widetilde{\mathbf{V}}_m$ can be obtained by checking the structures of the global variables listed in \eqref{g1}-\eqref{g3} and local variables listed in \eqref{e1}-\eqref{v1} \cite{6824195}. For simplicity, we omit the detailed expressions here.
\par
\begin{figure*}[b]
\hrule
\setcounter{equation}{77}
\begin{align} \label{wide_R}
& R_{m,k}\ui \geq \log_2 \left( 1 + \frac{ \mathrm{Tr}\left( \mathbf{H}_{m,m,k} \mathbf{W}_{m,k}\ui \right) }{\underset{k' \in \mathcal{K} \setminus \{k\} }{\sum} \hspace{-3mm}  \mathrm{Tr}\left( \mathbf{H}_{m,m,k} \mathbf{W}_{m,k'}\ui \right) + \mathbf{h}_{m,m,k}^H \mathbf{R}_{m}\ui \mathbf{h}_{m,m,k} + \hspace{-5mm}\underset{m' \in \mathcal{M} \setminus \{m\}}{\sum} \hspace{-5mm} e_{m',m,k}\ui + \hspace{-3mm} \underset{m'\in \mathcal{M}\setminus \{m\}}{\sum} \hspace{-2mm} t_{m',m,k}\ui + \sigma_{m,k}^2} \right) \notag \\
& \hspace{8mm} \overset{\triangle}{=} \widetilde{R}_{m,k}^{(i)}\left( \mathbf{W}^{(i)}_{m,k}, \mathbf{R}^{(i)}_m, \left\{ e_{m',m,k}^{(i)} \right\}_{\forall m' \neq m}, \left\{ t_{m',m,k}^{(i)} \right\}_{\forall m' \neq m} \right). \\
\setcounter{equation}{84}
& \bm{\Psi}_{m,k}^{(i)}\left( \mathbf{W}^{(i)}_{m,k}, \mathbf{R}^{(i)}_m, \psi_{m,k}^{(i)}, \xi_{m,k}^{(i)}, \left\{ u_{m'}^{(i)} \right\}_{m' \neq m} \right) \overset{\triangle}{=} 
\begin{bmatrix}
\psi_{m,k}^{(i)} \mathbf{I}_{N} \hspace{-5mm} & \hspace{-5mm} \mathbf{0} \\
\mathbf{0} \hspace{-5mm} & \hspace{-5mm} -\psi_{m,k}^{(i)} \left( \beta_{\mathbf{g}_m}^{(i)} \right)^2 + \xi_{m,k}^{(i)} \hspace{-3mm} \underset{m' \in \mathcal{M} \setminus \{m\}}{\sum} \hspace{-4mm} u_{m'}^{(i)} + \xi_{m,k}^{(i)} \sigma^2_{\text{e}} 
\end{bmatrix} -
\mathbf{D}_m^H
\widetilde{\mathbf{M}}^{(i)}_{m,k}
\mathbf{D}_m. \label{Psi_eq} \\
& \bm{\Phi}_m^{(i)} \left( \mathbf{R}^{(i)}_m, \lambda_m^{(i)}, u_m^{(i)} \right) \overset{\triangle}{=} \begin{bmatrix}
\lambda_m^{(i)} \mathbf{I}_{N} & \mathbf{0} \\
\mathbf{0} & -\lambda_m^{(i)} \left( \beta_{\mathbf{g}_m}^{(i)} \right)^2 - u_m^{(i)} 
\end{bmatrix}
+
\mathbf{D}_m^H
\mathbf{R}^{(i)}_m
\mathbf{D}_m. \label{Phi_eq} \\
\setcounter{equation}{95}
& \Pi_m \left( \mathbf{W}_{m,k}^{(1)}, \mathbf{R}_m^{(1)}, \left\{\mathbf{V}_{m'}\right\}_{\forall m' \neq m}, \delta_m \right) \overset{\triangle}{=} \mathrm{Tr}\left( \left( \overline{\mathbf{F}}_m + \hspace{-3mm} \underset{m' \in \mathcal{M} \setminus \{m\}}{\sum} \hspace{-3mm} \mathbf{V}_{m'} \right)^{-1} \right) - 2 a_m^2 ( \ul{\delta}_m - b_m )(\delta_m - \ul{\delta}_m) - a_m^2 ( \ul{\delta}_m - b_m )^2. \label{Pi}
\end{align}
\setcounter{equation}{108}
\end{figure*}
\par
The local variables of the $m$-th BS are collected by the set $\mathcal{V}_m$, given by
\begin{align}
\mathcal{V}_m \overset{\triangle}{=} \bigg\{ & \left\{\mathbf{W}^{(i)}_{m,k}\right\}_{\forall k}, \mathbf{R}^{(i)}_m, \left\{\xi_{m,k}^{(i)}\right\}_{\forall k}, \left\{\psi_{m,k}^{(i)}\right\}_{\forall k}, \lambda_m^{(i)}, \delta_m, \notag \\
& \hspace{2mm} \mathbf{e}_m^{(i)}, \mathbf{t}_m^{(i)}, \mathbf{u}_m^{(i)}, \widehat{\mathbf{V}}_m \bigg\}.
\end{align}
Substituting \eqref{e1} to \eqref{v1} into the constraints of problem \eqref{prob_1}, the feasible set can be decomposed into $M$ independent sets as follows
\begin{align}
\mathcal{C}_m \overset{\triangle}{=} & \Big\{ \mathcal{V}_m:
\mbox{C1} \mbox{:} \hspace{1mm} P_m \leq P_{\text{max}}, \notag \\
\widetilde{\mbox{C2}} \mbox{:} \hspace{1mm} & \underset{i \in \mathcal{I}}{\sum} \tau_i \widetilde{R}_{m,k}^{(i)}\left( \mathbf{W}^{(i)}_{m,k}, \mathbf{R}^{(i)}_m, \mathbf{e}_m\ui, \mathbf{t}_m\ui \right) \geq R_{\text{info},m,k}, \forall k, \notag \\
\widetilde{\mbox{C3}} \mbox{:} \hspace{1mm} &
\underset{i \in \mathcal{I}}{\sum} \tau_i \widetilde{\widehat{R}}_{m,k}\ui \leq R_{\text{leak},m,k}, \forall k, \notag \\
\widetilde{\mbox{C5a}} \mbox{:} \hspace{1mm} & \bm{\Psi}_{m,k}^{(1)}\left( \mathbf{W}^{(1)}_{m,k}, \mathbf{R}^{(1)}_m, \psi_{m,k}^{(1)}, \xi_{m,k}^{(1)}, \mathbf{u}_m^{(1)} \right) \succeq \mathbf{0}, \forall k, \notag \\
\mywidetilde{\mbox{C5b}} \mbox{:} \hspace{1mm} & \bm{\Psi}_{m,k}^{(2)}\left( \mathbf{W}^{(2)}_{m,k}, \mathbf{R}^{(2)}_m, \psi_{m,k}^{(2)}, \xi_{m,k}^{(2)}, \mathbf{u}_m^{(2)}, \delta_m \right) \succeq \mathbf{0}, \forall k, \notag \\
\mbox{C8:} \hspace{1mm} & e_{m,m',k}^{(i)} \geq \underset{k' \in \mathcal{K} }{\sum} \mathrm{Tr}\left( \mathbf{H}_{m,m',k} \mathbf{W}_{m,k'}^{(i)} \right), \hspace{1mm} \forall m' \neq m, \forall k, \notag \\
\mbox{C9:} \hspace{1mm} & t_{m,m',k}^{(i)} \geq \mathbf{h}_{m,m',k}^H \mathbf{R}_{m}^{(i)} \mathbf{h}_{m,m',k}, \forall m' \neq m, \forall k, \notag \\
\widetilde{\mbox{C10a}} \mbox{:} \hspace{1mm} & \bm{\Phi}_m^{(1)} \left( \mathbf{R}^{(1)}_m, \lambda_m^{(1)}, \mathbf{u}_m^{(1)} \right)  \succeq \mathbf{0}, \notag \\
\mywidetilde{\mbox{C10b}} \mbox{:} \hspace{1mm} & \bm{\Phi}_m^{(2)} \left( \mathbf{R}^{(2)}_m, \lambda_m^{(2)}, \mathbf{u}_m^{(2)}, \delta_m \right)  \succeq \mathbf{0}, \notag \\
\mywidetilde{\mbox{C11}} \mbox{:} \hspace{1mm} & \Pi_m\left( \mathbf{W}_{m,k}^{(1)}, \mathbf{R}_m^{(1)}, \widehat{\mathbf{V}}_{m}, \delta_m \right) \leq 0, \notag \\
\mbox{C12:} \hspace{1mm} & \overline{\mathbf{F}}_m \succeq \mathbf{V}_m
\Big\}.
\end{align}
As a result, problem \eqref{prob_1} is now recast as follows 
\begin{align} \label{before_admm}
\underset{\substack{ \mathcal{V}_m, \mathcal{G} }}{\mino} & \hspace*{3mm} \underset{m \in \mathcal{M}}{\sum} P_m + \underset{m \in \mathcal{M}}{\sum} \mathbb{I}_{\mathcal{C}_m} (\mathcal{V}_m) \notag\\
\subto & \hspace*{3mm} \mbox{C13a-C13d},
\end{align}
where $\mathbb{I}_{\mathcal{C}_m} (\mathcal{V}_m)$ is the indicator function defined as 
\begin{align}
\mathbb{I}_{\mathcal{C}_m}(\mathcal{V}_m) \overset{\triangle}{=}
\begin{cases}
0 & \hspace{-3mm}: \mathcal{V}_m \in \mathcal{C}_m, \\
\infty & \hspace{-3mm}: \text{otherwise},
\end{cases}
\end{align}
which is subsequently handled by enforcing the constraints specified in $\mathcal{C}_m$.
The augmented Lagrangian function of problem \eqref{before_admm} is given by
\begin{align}
& \mathcal{L}\left( \mathcal{V}_m, \mathcal{G}, \bm{\Theta} \right) = \underset{m \in \mathcal{M}}{\sum} P_m + \underset{m \in \mathcal{M}}{\sum} \mathbb{I}_{\mathcal{C}_m} (\mathcal{V}_m) + \underset{m \in \mathcal{M}}{\sum} h_m,
\end{align}
where
\begin{align}
& h_m = \underset{i \in \mathcal{I}}{\sum} \Bigg( \frac{\rho_1}{2} \left\| \mathbf{e}_m^{(i)} - \mathbf{E}_m \mathbf{e}^{(i)} + \frac{\bm{\lambda}_{1,m}^{(i)}}{\rho_1} \right\|^2 \notag \\
& + \frac{\rho_2}{2} \left\| \mathbf{t}_m^{(i)} \hspace{-1mm} - \hspace{-1mm} \mathbf{T}_m \mathbf{t}^{(i)} \hspace{-1mm} + \hspace{-1mm} \frac{\bm{\lambda}_{2,m}^{(i)}}{\rho_2} \right\|^2 \hspace{-2mm} + \hspace{-1mm} \frac{\rho_3}{2} \left\| \mathbf{u}_m^{(i)} \hspace{-1mm} - \hspace{-1mm} \mathbf{U}_m \mathbf{u}^{(i)} \hspace{-1mm} + \hspace{-1mm} \frac{\bm{\lambda}_{3,m}^{(i)}}{\rho_3} \right\|^2 \Bigg) \notag \\
& + \frac{\rho_4}{2} \left\| \widehat{\mathbf{V}}_m - \widetilde{\mathbf{V}}_m \mathbf{V} + \frac{\bm{\Lambda}_{4,m}}{\rho_4} \right\|^2,
\end{align}
Here, $\bm{\lambda}_{1,m}^{(i)}$, $\bm{\lambda}_{2,m}^{(i)}$, $\bm{\lambda}_{3,m}^{(i)}$, and $\bm{\Lambda}_{4,m}$ are dual variables with respect to constraints \mbox{C13a-d}, respectively, and are collected by the set $\bm{\Theta}$. Scalars $\rho_1$, $\rho_2$, $\rho_3$, and $\rho_4$ are corresponding penalty parameters. According to the principle of ADMM, for given primal variables, the dual variables are updated by
\begin{align}
\bm{\lambda}_{1,m}^{(i)} & = \underline{\bm{\lambda}}_{1,m}^{(i)} + \rho_1 \left( \mathbf{e}_m^{(i)} - \mathbf{E}_m \mathbf{e}^{(i)} \right), \label{dual_1}\\
\bm{\lambda}_{2,m}^{(i)} & = \underline{\bm{\lambda}}_{2,m}^{(i)} + \rho_2 \left( \mathbf{t}_m^{(i)} - \mathbf{T}_m \mathbf{t}^{(i)} \right), \label{dual_2}\\
\bm{\lambda}_{3,m}^{(i)} & = \underline{\bm{\lambda}}_{3,m}^{(i)} + \rho_3 \left( \mathbf{u}_m^{(i)} - \mathbf{U}_m \mathbf{u}^{(i)} \right), \label{dual_3}\\
\bm{\Lambda}_{4,m} & = \underline{\bm{\Lambda}}_{4,m} + \rho_4 \left( \widehat{\mathbf{V}}_m - \widetilde{\mathbf{V}}_m \mathbf{V} \right), \label{dual_4}
\end{align}
where $\ul{\bm{\lambda}}_{1,m}^{(i)}$, $\ul{\bm{\lambda}}_{2,m}^{(i)}$, $\ul{\bm{\lambda}}_{3,m}^{(i)}$, and $\ul{\bm{\Lambda}}_{4,m}$ denote the solutions obtained in the previous iteration. The primal variables are updated by alternatingly minimizing the augmented Lagrangian function $\mathcal{L}\left( \mathcal{V}_m, \mathcal{G}, \bm{\Theta} \right)$.
For given $\ul{\mathcal{G}}$ and $\underline{\bm{\Theta}}$, the local variables are updated by solving
\begin{align} \label{local_prob}
\underset{\substack{ \mathcal{V}_m }}{\mino} & \hspace*{3mm} \mathcal{L}\left( \mathcal{V}_m, \ul{\mathcal{G}}, \underline{\bm{\Theta}} \right),
\end{align}
which can be decomposed into $M$ independent problems for decentralized optimization. For given $\ul{\mathcal{V}}_m$ and $\underline{\bm{\Theta}}$, the global variables are updated by solving
\begin{align} \label{global_prob}
\underset{\substack{ \mathcal{G} }}{\mino} & \hspace*{3mm} \mathcal{L}\left( \ul{\mathcal{V}}_m, \mathcal{G}, \underline{\bm{\Theta}} \right).
\end{align}
Next, we update $\ul{\mathcal{V}}_m$ and $\ul{\mathcal{G}}$ by alternatingly minimizing $\mathcal{L}\left( \ul{\mathcal{V}}_m, \mathcal{G}, \underline{\bm{\Theta}} \right)$.

\subsection{Update Local Variables $\mathcal{V}_m$}
Problem \eqref{local_prob} can be decomposed into $M$ independent problems given by
\begin{equation}
\begin{aligned}
\underset{\substack{ \mathcal{V}_m}}{\mino} & \hspace*{3mm} P_m + h_m, \\
\subto & \hspace*{3mm} \mathcal{V}_m \in \mathcal{C}_m,
\end{aligned}
\hspace{5mm} \forall m \in \mathcal{M},
\end{equation}
where the non-convexity comes from the variable coupling issue. To handle this, we divide the variables into two sets, which are given by, respectively,
\begin{align}
\mathcal{V}_{1,m} \overset{\triangle}{=} \bigg\{ & \left\{\mathbf{W}^{(i)}_{m,k}\right\}_{\forall k}, \mathbf{R}^{(i)}_m, \delta_m, \mathbf{e}_m^{(i)}, \mathbf{t}_m^{(i)}, \mathbf{u}_m^{(i)}, \widehat{\mathbf{V}}_m \bigg\}, \\
\mathcal{V}_{2,m} \overset{\triangle}{=} \bigg\{ & \left\{\xi_{m,k}^{(i)}\right\}_{\forall k}, \left\{\psi_{m,k}^{(i)}\right\}_{\forall k}, \lambda_m^{(i)} \bigg\}.
\end{align}
The subproblem with respect to $\mathcal{V}_{1,m}$ is given by
\begin{align} \label{local_prob_1}
\underset{\substack{ \mathcal{V}_{1,m}}}{\mino} & \hspace*{3mm} P_m + h_m \notag \\
\subto & \hspace*{3mm} \mathcal{V}_{1,m} \in \mathcal{C}_{1,m},
\end{align}
where
\begin{align}
\mathcal{C}_{1,m} \overset{\triangle}{=} \big\{ \mathcal{V}_{1,m}: & \mbox{C1}, \widetilde{\mbox{C2}}, \widetilde{\mbox{C5a}}, \mywidetilde{\mbox{C5b}}, \mbox{C8}, \mbox{C9}, \widetilde{\mbox{C10a}}, \mywidetilde{\mbox{C10b}}, \notag \\
& \mywidetilde{\mbox{C11}}, \mbox{C12} \big\}.
\end{align}
To tackle the non-convex constraint $\widetilde{\mbox{C2}}$, we first rewrite $\widetilde{R}_{m,k}^{(i)}$ as
\begin{align}
\widetilde{R}_{m,k}^{(i)} = \widetilde{X}_{m,k}^{(i)} - \widetilde{Y}_{m,k}^{(i)},
\end{align}
where
\begin{align}
\widetilde{X}_{m,k}^{(i)} & = \log_2\Big( \mathrm{Tr}\big( \mathbf{H}_{m,m,k} \mathbf{W}_{m,k}\ui \big) + \widetilde{Z}_{m,k}\ui \Big), \\
\widetilde{Y}_{m,k}^{(i)} & = \log_2 \widetilde{Z}_{m,k}\ui.
\end{align}
Here, $\widetilde{Z}_{m,k}\ui$ is given by
\begin{align}
\widetilde{Z}_{m,k}\ui = & \underset{k' \in \mathcal{K} \setminus \{k\} }{\sum} \hspace{-3mm}  \mathrm{Tr}\left( \mathbf{H}_{m,m,k} \mathbf{W}_{m,k'}\ui \right) + \mathbf{h}_{m,m,k}^H \mathbf{R}_{m}\ui \mathbf{h}_{m,m,k} \notag \\
& + \overline{e}_{m,k}\ui + \overline{t}_{m,k}\ui + \sigma_{m,k}^2.
\end{align}
Then, a convex upper bound of $\widetilde{Y}_{m,k}^{(i)}$ is constructed by \cite{11048972}
\begin{align}
& \widetilde{Y}_{m,k}^{(i)} \leq \ul{\widetilde{Y}}_{m,k}^{(i)} + \underset{\overline{k} \in \mathcal{K}}{\sum} \mathrm{Tr}\left( \nabla^H_{\mathbf{W}_{m,\overline{k}}^{(i)}} \ul{\widetilde{Y}}_{m,k}^{(i)} \left( \mathbf{W}_{m,\overline{k}}^{(i)} - \underline{\mathbf{W}}_{m,\overline{k}}^{(i)} \right) \right) \notag \\
& + \mathrm{Tr}\left( \nabla^H_{\mathbf{R}_m^{(i)}} \ul{\widetilde{Y}}_{m,k}^{(i)} \left( \mathbf{R}_m^{(i)} - \underline{\mathbf{R}}_m^{(i)} \right) \right) \notag \\
& + \nabla_{\overline{e}_{m,k}^{(i)}} \ul{\widetilde{Y}}_{m,k}^{(i)} \left( \overline{e}_{m,\overline{k}}^{(i)} - \ul{\overline{e}}_{m,k}^{(i)} \right) +  \nabla_{\overline{t}_{m,k}^{(i)}} \ul{\widetilde{Y}}_{m,k}^{(i)} \left( \overline{t}_{m,\overline{k}}^{(i)} - \ul{\overline{t}}_{m,k}^{(i)} \right) \notag \\
& \overset{\triangle}{=} \overline{\widetilde{Y}}_{m,k}^{(i)},
\end{align}
where $\ul{\widetilde{Y}}_{m,k}^{(i)}$ denotes the value of
$\widetilde{Y}_{m,k}^{(i)}$ evaluated by $\ul{\mathbf{W}}_{m,k}^{(i)}, \ul{\mathbf{R}}_m^{(i)}, \ul{\overline{e}}_{m,k}^{(i)}$, and $\ul{\overline{t}}_{m,k}^{(i)}$, which denote the solution obtained in the previous iteration. $\nabla_{\mathbf{W}_{m,\overline{k}}^{(i)}} Y_{m,k}^{(i)}$,
$\nabla_{\mathbf{R}_m^{(i)}} Y_{m,k}^{(i)}$, $\nabla_{\overline{e}_{m,k}^{(i)}} \widetilde{Y}_{m,k}^{(i)}$ and $\nabla_{\overline{t}_{m,k}^{(i)}} \widetilde{Y}_{m,k}^{(i)}$
denote the gradients of $Y_{m,k}^{(i)}$ with respect to $\mathbf{W}_{m,\overline{k}}^{(i)}$, $\mathbf{R}_m^{(i)}$, $\overline{e}_{m,k}^{(i)}$ and $\overline{t}_{m,k}^{(i)}$, which are given by, respectively,
\begin{align}
\nabla^H_{\mathbf{W}_{m,\overline{k}}^{(i)}} Y_{m,k}^{(i)} & = 
\begin{cases}
0 & \hspace{-3mm}: \overline{k} = k, \\
\frac{1}{\ln 2} \frac{\mathbf{H}_{m,m,k}}{Z_{m,k}\ui} & \hspace{-3mm}: \overline{k} \neq k,
\end{cases} \\
\nabla^H_{\mathbf{R}_m^{(i)}} Y_{m,k}^{(i)} & = \frac{1}{\ln 2} \frac{\mathbf{H}_{m,m,k}}{Z_{m,k}\ui}, \\
\nabla_{\overline{e}_{m,k}^{(i)}} \widetilde{Y}_{m,k}^{(i)} & = \frac{1}{\ln 2} \frac{1}{Z_{m,k}\ui}, \\ \nabla_{\overline{t}_{m,k}^{(i)}} \widetilde{Y}_{m,k}^{(i)} & = \frac{1}{\ln 2} \frac{1}{Z_{m,k}\ui}.
\end{align}
Then, constraint $\widetilde{\mbox{C2}}$ is transformed into
\begin{align}
\mywidetilde{\mbox{C2}} \mbox{:} \hspace{1mm} \underset{i \in \mathcal{I}}{\sum} \tau_i \big( X_{m,k}^{(i)} - \overline{\widetilde{Y}}_{m,k}^{(i)} \big) \geq R_{\text{info},m,k},\hspace*{1mm}\forall k.
\end{align}
By replacing constraint $\mywidetilde{\mbox{C2}}$ with constraint $\mbox{C2}$ in set $\mathcal{C}_{1,m}$, $\mathcal{C}_{1,m}$ is transformed into $\widetilde{\mathcal{C}}_{1,m}$,
Problem \eqref{local_prob_1} is then recast as the following convex optimization problem
\begin{align} \label{local_prob_2}
\underset{\substack{ \mathcal{V}_{1,m}}}{\mino} & \hspace*{3mm} P_m + h_m, \notag \\
\subto & \hspace*{3mm} \mathcal{V}_{1,m} \in \widetilde{\mathcal{C}}_{1,m},
\end{align}
which can be solved efficiently.
\par
The subproblem with respect to $\mathcal{V}_{2,m}$ is given by
\begin{align} \label{local_prob_3}
\underset{\substack{ }}{\mathrm{Find}} & \hspace*{3mm} \mathcal{V}_{2,m}, \notag \\
\subto & \hspace*{3mm} \mathcal{V}_{2,m} \in \mathcal{C}_{2,m},
\end{align}
where $\mathcal{C}_{2,m} \overset{\triangle}{=} \big\{ \mathcal{V}_{2,m}: \widetilde{\mbox{C3}}, \widetilde{\mbox{C5a}}, \mywidetilde{\mbox{C5b}}, \widetilde{\mbox{C10a}}, \mywidetilde{\mbox{C10b}} \big\}$. Similar to the steps for addressing problem \eqref{feasibilty_check}, we update $\mathcal{V}_{2,m}$ by solving a convex information leakage minimization problem, which is formulated as follows
\begin{align} \label{local_prob_4}
\underset{\substack{ \mathcal{V}_{2,m} }}{\mino} & \hspace*{3mm} \underset{i \in \mathcal{I}}{\sum} \tau_i \widetilde{\widehat{R}}_{m,k}\ui \notag \\
\subto & \hspace*{3mm} \mathcal{V}_{2,m} \in \mathcal{C}_{2,m} \setminus \{ \xi_{m,k}\ui: \widetilde{\mbox{C3}} \}.
\end{align}

\subsection{Update Global Variables $\mathcal{G}$}
Problem \eqref{global_prob} for updating $\mathcal{G}$ is equivalent to
\begin{align} \label{update_G_0}
\underset{\substack{ \mathcal{G} }}{\mino} & \hspace*{3mm}  \underset{m \in \mathcal{M}}{\sum} h_m.
\end{align}
The optimal $\mathbf{e}^{(i)}$ is updated by
\begin{align} \label{update_e}
\mathbf{e}^{(i)} & = \underset{\mathbf{e}^{(i)}}{\arg \min} \underset{m \in \mathcal{M}} {\sum} \left\| \mathbf{e}_m^{(i)} - \mathbf{E}_m \mathbf{e}^{(i)} + \frac{\bm{\lambda}_{1,m}^{(i)}}{\rho_1} \right\|^2 \notag \\
& = \left( \underset{m \in \mathcal{M}}{\sum} \mathbf{E}_m^T \mathbf{E}_m \right)^{-1} \left( \underset{m \in \mathcal{M}}{\sum} \mathbf{E}_m^T \left( \mathbf{e}_m^{(i)} + \frac{\bm{\lambda}_{1,m}^{(i)}}{\rho_1} \right) \right).
\end{align}
Similarly, the optimal $\mathbf{t}^{(i)}$, $\mathbf{u}^{(i)}$, and $\mathbf{V}$ are updated by
\begin{align}
& \mathbf{t}^{(i)} = \left( \underset{m \in \mathcal{M}}{\sum} \mathbf{T}_m^T \mathbf{T}_m \right)^{-1} \left( \underset{m \in \mathcal{M}}{\sum} \mathbf{T}_m^T \left( \mathbf{t}_m^{(i)} + \frac{\bm{\lambda}_{2,m}^{(i)}}{\rho_2} \right) \right), \label{update_t} \\
& \mathbf{u}^{(i)} = \left( \underset{m \in \mathcal{M}}{\sum} \mathbf{U}_m^T \mathbf{U}_m \right)^{-1} \left( \underset{m \in \mathcal{M}}{\sum} \mathbf{U}_m^T \left( \mathbf{u}_m^{(i)} + \frac{\bm{\lambda}_{3,m}^{(i)}}{\rho_3} \right) \right), \label{update_u} \\
& \mathbf{V} = \left( \underset{m \in \mathcal{M}}{\sum} \widetilde{\mathbf{V}}_m^H \widetilde{\mathbf{V}}_m \right)^{-1} \left( \underset{m \in \mathcal{M}}{\sum} \widetilde{\mathbf{V}}_m^H \left( \widehat{\mathbf{V}}_m + \frac{\bm{\Lambda}_{4,m}}{\rho_4} \right) \right). \label{update_V}
\end{align}
\par
The ADMM-based decentralized design for solving problem \eqref{main_prob} is summarized in $\textbf{Algorithm 2}$. $f_2^{(n)}$ denotes the objective function value of \eqref{main_prob} at the $n$-th iteration. The convergence and computational complexity of Algorithm 2 are discussed as follows.
\begin{algorithm}[t]
\caption{ADMM-based Decentralized Algorithm for Solving Problem \eqref{main_prob}}
\begin{algorithmic}[1]
\small
\STATE Set iteration index $n=1$, scaling factor for the penalty parameters $\varsigma > 1$, error tolerance $\epsilon_2$.
\REPEAT
\STATE Update $\mathcal{V}_{1,m}$ by solving problem \eqref{local_prob_2} at each BS
\STATE Update $\mathcal{V}_{2,m}$ by solving problem \eqref{local_prob_4} at each BS
\STATE Update $\mathcal{G}$ by \eqref{update_e}, \eqref{update_t}, \eqref{update_u} and \eqref{update_V} at central controller
\STATE Update $\bm{\lambda}_{1,m}\ui$, $\bm{\lambda}_{2,m}\ui$, $\bm{\lambda}_{3,m}\ui$, $\bm{\Lambda}_{4,m}$ by \eqref{dual_1}, \eqref{dual_2}, \eqref{dual_3} and \eqref{dual_4}, respectively, at each BS
\STATE Update penalty parameters $\rho_1 \leftarrow \varsigma \rho_1$, $\rho_2 \leftarrow \varsigma \rho_2$, $\rho_3 \leftarrow \varsigma \rho_3$, $\rho_4 \leftarrow \varsigma \rho_4$
\STATE Set $n \leftarrow n + 1$
\UNTIL $\frac{\left|f_2^{(n)} - f_2^{(n-1)}\right|}{f_2^{(n-1)}} \leq \epsilon_2$
\end{algorithmic}
\end{algorithm}
\par
The objective function value of problem \eqref{local_prob_2} is an upper bound of the objective function value of problem \eqref{local_prob_1}. For any fixed penalty parameters, this upper bound is monotonically tightened at each iteration. Together by alternatingly solving \eqref{local_prob_4}, and \eqref{update_G_0}, the objective function value of problem \eqref{main_prob} is non-increasing \cite{xu2025joint}. According to \cite{boyd2011distributed}, the proposed algorithm is guaranteed to converge to a stationary point of problem \eqref{main_prob}. The computational complexity is dominated by solving problem \eqref{local_prob_2} at each BS, given by
$\mathcal{O} \big( \log\frac{1}{\varepsilon} (KN^3 + K^2N^2 +K^3) \big)$.
\begin{remark}
The decentralized design significantly reduces computational complexity and information exchange overhead compared to the centralized design. Specifically, the decentralized design enables the small-scale subproblems to be computed in parallel at each BS. Hence, the computational complexity remains independent of the number of BSs, $M$, whereas that of the centralized design scales proportionally to $M^4$.
\par
From the perspective of information exchange overhead, the centralized design requires the transmission of the received echoes and the CSI of all BSs, with an information exchange overhead of $\mathcal{O}(MNL+M^2KN)$.
In contrast, the decentralized design involves only sharing estimated sensing results, the global and local variables for optimization, with an information exchange overhead of $\mathcal{O}(M^2KI_{\text{iter}})$, where $I_{\text{iter}}$ denotes the iteration number. Here, we note that the number of time slots accumulated for sensing, i.e., $L$, is generally significantly larger than $M$, $K$, and $I_{\text{iter}}$ \cite{10364735}. Hence, the information exchange overhead of decentralized design is largely reduced.
\end{remark}
\section{Simulation Results}
In this section, the effectiveness of the proposed algorithm design is validated by simulation results. We consider a networked ISAC system with $M=3$ BSs. Each BS is equipped with $N = 4$ antennas and serves $K=2$ legitimate users. The BSs are located at $[15, 22.5]$, $[-25, 25]$, and $[15, -30]$, where the unit is meter. The eavesdropper is located at the origin. The system operates at a frequency of $5$ GHZ. The duration ratios for the two stages are given by $\tau_1 = 0.2$ and $\tau_2 = 0.8$, respectively.\footnote{The system performance relies on the duration ratios of the two stages, which can be optimally determined using the line search method \cite{10349846}. In this paper, we focus on the adaptive sensing performance design with fixed duration ratios of the two stages. } The achievable rate requirement of users is set to $R_{\text{info},m,k} = 5$ bits/s/Hz, and the maximum tolerable information leakage rate is set to $R_{\text{leak},m,k} = 0.5$ bits/s/Hz. The maximum transmit power at each BS is given by $P_{\text{max}} = 20$ dBm. The noise power at the users, the eavesdropper, and the sensing receiver are set to $\sigma_{m,k}^2 = \sigma_{\text{e}}^2 = \sigma_{\text{s}}^2 = -100$ dBm. The number of samples used for sensing is set to $L = 1024$ \cite{10364735}. The penalty parameters are initialized as $\rho_i = 10^4$, $i = 1,2,3,4$, and the scaling factor is set to $\varsigma = 1.5$.
\par
For comparison purposes, we consider the following baseline schemes in the existing literature.
\begin{itemize}
\item \textbf{Baseline scheme 1:} This baseline scheme follows the separated two-stage sensing-enhanced secure communication design \cite{9645576}. Specifically, in the first stage, the system aims to achieve the best sensing performance. Then, in the second stage, the system performs secure communication based on the sensing results obtained in the first stage. The baseline scheme uses a centralized design.
\item \textbf{Baseline scheme 2:} This baseline scheme adopts the fixed sensing performance for secure communication \cite{11126093}. The transmit beamforming is designed to satisfy the fixed sensing performance requirement and secure communication requirements. The baseline scheme uses a decentralized design.
\end{itemize}
We refer to the proposed centralized and decentralized designs as ``Proposed Scheme 1" and ``Proposed Scheme 2", respectively.
\begin{figure}[t]
\centering
\includegraphics[width=3.4in]{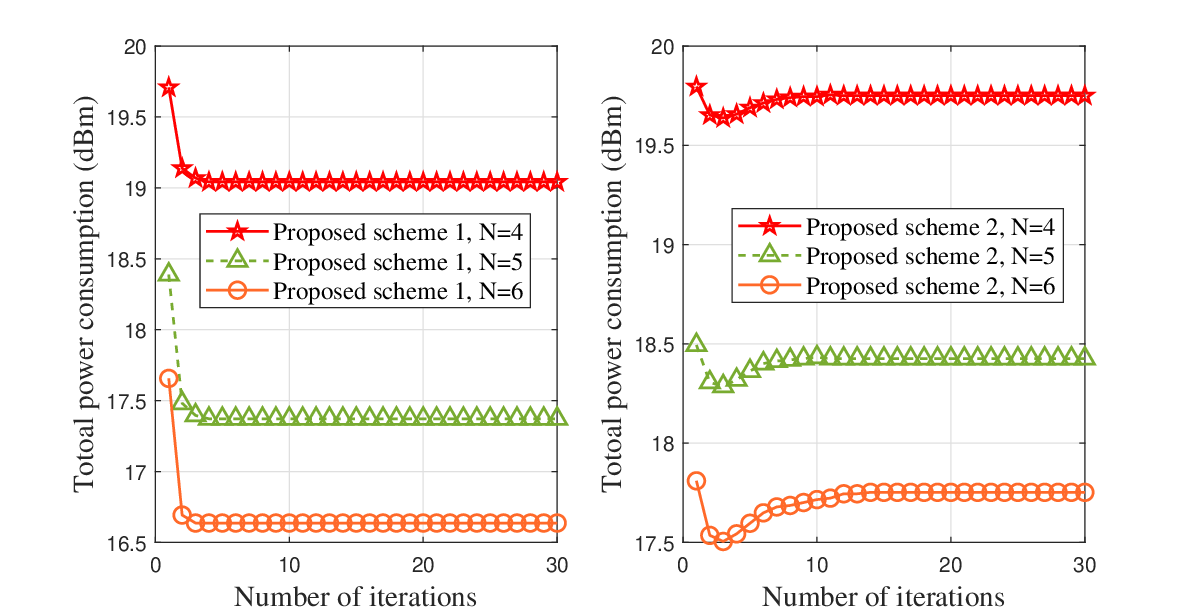}
\caption{Convergence behavior of Algorithm 1 (Left) and Algorithm 2 (Right).}
\label{figure:convergence}
\end{figure}
\begin{figure}[t]
\centering
\includegraphics[width=3.4in]{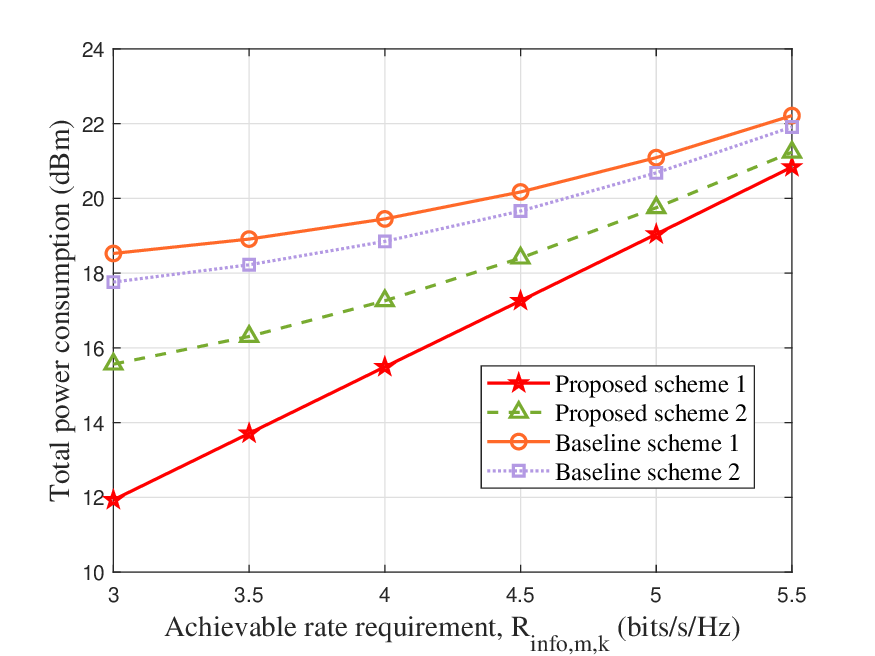}
\caption{Total power consumption (dBm) versus achievable rate requirement $R_{\text{info},m,k}$ (bits/s/Hz).}
\label{figure:p_info}
\end{figure}
\subsection{Algorithm Convergence}
Fig. \ref{figure:convergence} shows the convergence behavior of the proposed algorithms with different numbers of antennas at each BS. As can be observed, the total power consumption of Algorithm 1 monotonically decreases. For Algorithm 2, the total power consumption initially decreases and then increases until convergence. This phenomenon is caused by increasingly larger penalty parameters at each step. Specifically, the initial penalty parameters are not too large, allowing for constraint violations to occur. Hence, the equality constraints are relaxed, and the objective function value decreases. As the penalty parameters increase, the equality constraints are gradually enforced, resulting in a progressively shrinking feasible region. Hence, the objective function value increases until convergence. Furthermore, it can be observed that the objective function value decreases with the number of antennas at each BS for both algorithms. 
\subsection{Performance Evaluation}
\subsubsection{Achievable Rate Requirement}
Fig. \ref{figure:p_info} shows the total power consumption (dBm) versus the achievable rate requirement $R_{\text{info},m,k}$ (bits/s/Hz). As can be observed, both proposed schemes outperform the baseline schemes, and proposed scheme 1 achieves greater power savings than proposed scheme 2, showing the performance benefits of the centralized design relative to the decentralized design. The performance gap between baseline scheme 1 and the proposed schemes arises because baseline scheme 1 separately designs the two stages, resulting in excessive power in the first stage to minimize sensing error. The performance improvement demonstrates the advantage of the proposed joint two-stage design, where power can be more effectively allocated among the two stages to enhance system performance.
Compared to baseline scheme 2, which adopts fixed sensing performance, the proposed schemes can adaptively determine the sensing performance to balance the power consumption between sensing in the first stage and secure communication in the second stage.
\begin{figure}[t]
\centering
\includegraphics[width=3.4in]{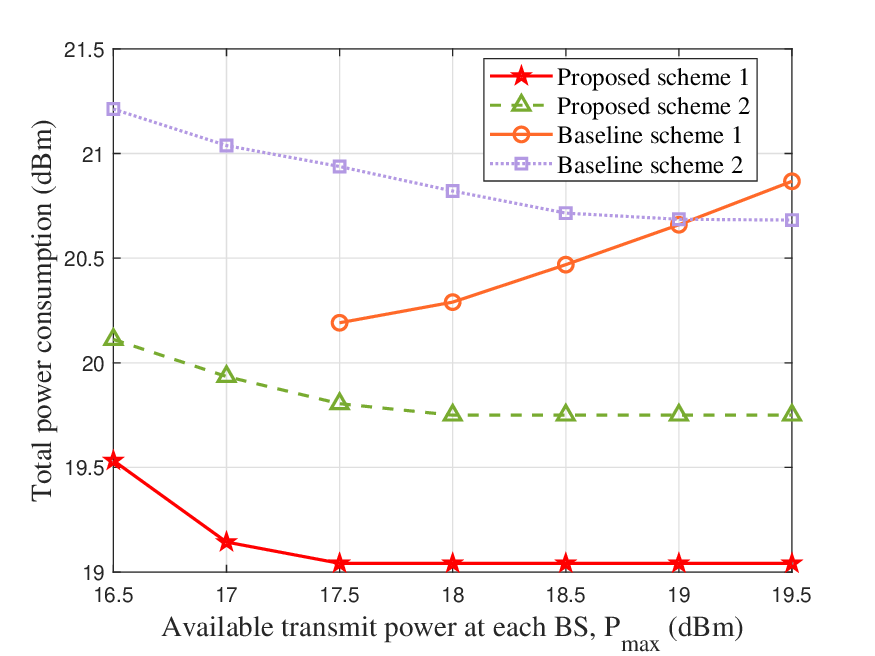}
\caption{Total power consumption (dBm) versus available transmit power at each BS $P_{\text{max}}$ (dBm).}
\label{figure:p_pmax}
\end{figure}
\subsubsection{Available Transmit Power at Each BS}
We investigate the impact of the available transmit power $P_{\text{max}}$ in Fig. \ref{figure:p_pmax}. It is observed that the total power consumption decreases with $P_{\text{max}}$ for the proposed schemes and baseline scheme 2, whereas baseline scheme 1 shows an increase in power consumption. This is because, for the proposed schemes and baseline scheme 2, a larger $P_{\text{max}}$ enlarges the feasible region of the optimization problem via constraint C1. This allows for more flexible power allocations among the BSs for collaboration, thereby leading to better performance.
However, for baseline scheme 1, the power in the first stage is dedicated to sensing, resulting in excellent sensing performance. Further increasing power yields limited improvements in sensing performance, offering minimal benefits for secure communication in the second stage. These benefits fail to offset the additional power consumption required for sensing in the first stage. Furthermore, for $P_{\text{max}} < 17.5$ dBm, the resulting sensing performance in the first stage causes infeasibility in the second stage of baseline scheme 1.


\begin{figure}[t]
\centering
\includegraphics[width=3.4in]{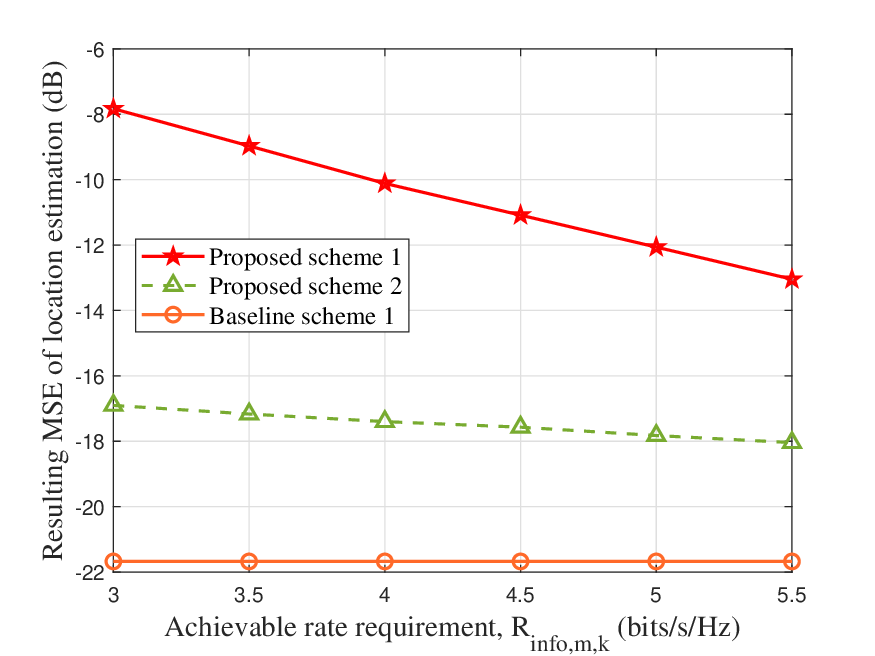}
\caption{Adaptive MSE of location estimation (dB) versus achievable rate requirement $R_{\text{info},m,k}$ (bits/s/Hz).}
\label{figure:mse_info}
\end{figure}
\subsection{Adaptive Sensing Performance}
\subsubsection{MSE of Location Estimation versus Achievable Rate Requirement}
Fig. \ref{figure:mse_info} shows the MSE of location estimation (dB) versus achievable rate requirement $R_{\text{info},m,k}$ (bits/s/Hz). The baseline scheme 2 is omitted in the figure because its sensing performance is predetermined. As shown in the figure, the sensing performance of the proposed schemes gradually decreases to adapt to the increasing achievable rate requirement. This is because, under the constraint of the information leakage rate, achieving a higher achievable rate requires more precise sensing performance to reduce the channel uncertainty. 
The proposed schemes benefit from the adaptation of sensing performance to achieve a favorable system performance, validating the effectiveness of the adaptive sensing performance.
Besides, it is worth noting that higher sensing performance is not always beneficial. Baseline scheme 1 achieves the best sensing performance but exhibits inferior performance, as shown in Fig. \ref{figure:p_info} and Fig. \ref{figure:p_pmax}. 
\begin{figure}[t]
\centering
\includegraphics[width=3.4in]{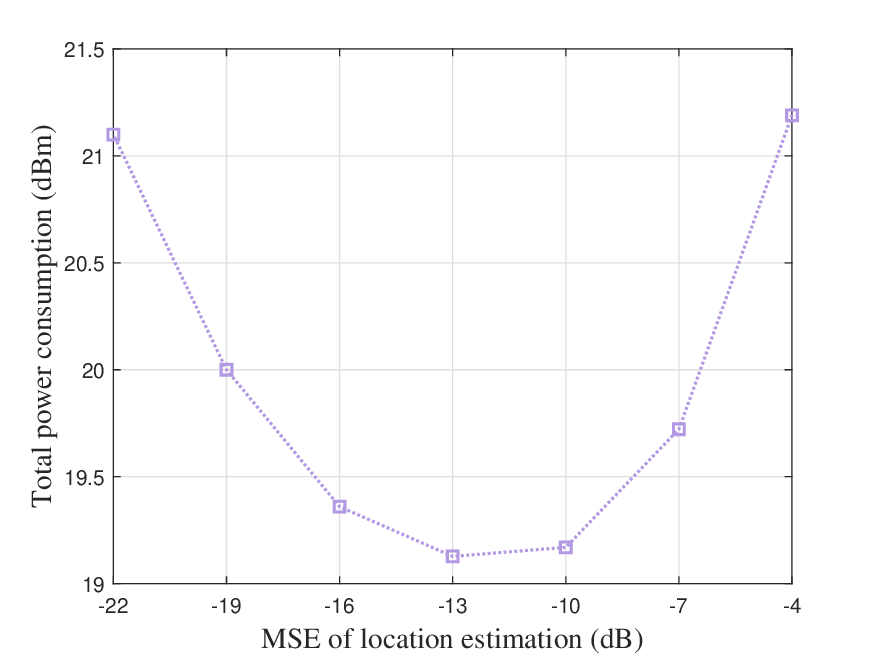}
\caption{Total power consumption (dBm) versus MSE of location estimation (dB).}
\label{figure:p_mse}
\end{figure}
\subsubsection{Optimal Sensing Performance}
Fig. \ref{figure:p_mse} shows the performance of the decentralized design under different sensing performance levels.
It is observed that system performance reaches its peak at a specific sensing performance point, characterized by the optimal trade-off between sensing accuracy and secure communication efficiency. Specifically, higher sensing performance requirements lead to significant energy consumption, which far exceeds the power savings brought by the improved sensing capabilities. Conversely, less sensing performance requirements result in significant channel uncertainty for the eavesdropper, leading to considerably high energy expenditure for secure communication. However, in practice, the search range for sensing performance is unknown, and determining the optimal sensing performance requires exhaustive search, which hinders the real-time deployment of the algorithm. Therefore, enforcing explicit sensing performance requirements is unnecessary and counterproductive.


\section{Conclusion}
In this paper, we have explored sensing-enhanced secure communication in networked ISAC systems with adaptive sensing performance. In particular, we have optimized the sensing and communication signals to minimize total power consumption. We have implicitly integrated the sensing performance into the information leakage rate, allowing it to be adaptively optimized for maximizing overall system performance. We have considered both centralized and decentralized schemes, where we have proposed a BCD-based framework to tackle the unique challenge of centralized design, and the challenge of decentralized design has been solved by the consensus ADMM method.
Simulation results have demonstrated the advantages of the proposed design by adapting the sensing performance according to varying communication requirements, fully leveraging the synergy between sensing and communication.
This indicates that explicit sensing requirements are unnecessary and potentially harmful.

\bibliographystyle{IEEEtran}
\bibliography{Reference_List}
\end{document}